\documentclass[12pt,a4paper]{article}
\usepackage[utf8]{inputenc}
\usepackage[english]{babel}
\usepackage{amsmath}
\usepackage{amsfonts}
\usepackage{amssymb}
\usepackage{graphicx}
\usepackage{cancel}
\usepackage{color}
\usepackage{slashed, epsf, latexsym}
\usepackage{epsfig}
\usepackage{graphics}
\usepackage{jheppub}

\begin{document}
\preprint{}
	\title{A leading-order comparison between fluid-gravity and membrane-gravity dualities}
	\author[]{Sayantani Bhattacharyya} 
	\author[]{, Parthajit Biswas}
	\author[]{and Milan Patra}
	\affiliation[]{National Institute of Science Education and Research, HBNI, Bhubaneshwar 752050, Odisha, India}
	\emailAdd{sayanta@niser.ac.in}
	\emailAdd{parthajit.biswas@niser.ac.in}
	\emailAdd{milan.patra@niser.ac.in}

\abstract{In this note,  we have compared  two different perturbation techniques that are used to generate dynamical black-brane solutions to Einstein's equations in presence of negative cosmological constant. One is the `derivative expansion', where the gravity solutions are in one-to-one correspondence with the solutions of relativistic Navier-Stokes equation. The second is the expansion in terms of inverse power of space-time dimensions and here the gravity solutions are dual to a co-dimension one dynamical membrane, embedded in AdS space and coupled to a velocity field. We have shown that in large number of space-time dimensions, there exists an overlap regime between these two perturbation techniques and we matched the two gravity solutions along with their dual systems upto the first non-trivial order in the expansion parameter on both sides. In the process,  we established a one-to-one map between dynamical black-brane geometry and the AdS space, which exists even when the number of dimensions is finite. }

\maketitle

\section{Introduction:}\label{sec:intro}
It is very hard to solve Einstein's equations - the key equation governing the dynamics of space-time, even at classical level. Only few exact solutions are known, mostly being static or stationary.  To handle non-trivial dynamics,  we have to take recourse of perturbation. \\
 Two such important perturbation schemes, which can handle dynamical fluctuations around static solutions even at non-linear level, are `derivative expansion'  \cite{Bhattacharyya:2007vs,nonlinfluid,Bhattacharyya:2008xc,Haack:2008cp,arbitrarydim,Bhattacharyya:ins} and expansion in inverse powers of dimension \cite{arbBack,membrane,yogesh1,yogesh2,Emparan:2013moa,Emparan:2013xia,Emparan:2014cia,EmparanCoupling,QNM:Emparan,Dandekar:2017aiv}. The first one generates `black-hole' type solutions (i.e., space-time with singularity shielded behind the horizon) that are in one to one correspondence with the  solutions of relativistic Navier Stokes equation\footnote{See \cite{Hubeny:2011hd}  and  references therein}  whereas the second one generates similar `black hole type' solutions, but dual to the dynamics of a codimension one membrane embedded in the asymptotic geometry\footnote{See \cite{secondorder} and references therein}.\\
It is  natural to ask whether it is possible to apply both the perturbation techniques simultaneously in any regime(s) of the parameter-space of the solutions, and if so, how the two solutions compare in those regimes.  In this note we would like to answer these two questions. In a nutshell our final result  is only what is expected.
\begin{itemize}
\item  It is possible to apply both the perturbation techniques simultaneously. 
Further, in the regime where both  $D$ is large and derivatives are small in an appropriate sense, we could treat $\left(1\over D\right)$ and $\partial_\mu$ (with respect to some length scale) as two independent small parameters, with no constraint on their  ratio.
\item In other words, if the metric dual to hydrodynamics is further expanded in inverse powers of dimension, it matches with the metric dual to membrane-dynamics, again expanded in terms of derivatives.
\end{itemize}
However, this matching is not at all manifest. We could see it only after some appropriate gauge or coordinate transformation of one solution to the other. The whole subtlety of our computation lies in finding the appropriate coordinate transformation. \\
The `large-$D$' expansion technique, as described in \cite{arbBack}, generates the dynamical black brane geometry  in  a `split form'  where the  full metric could always be written as a sum of pure AdS metric and something else. In other words, the black-brane space-time, constructed through `large-$D$' approximation would always  admit a very particular point-wise map  to   pure AdS geometry.\\ 
On the other hand, the space-time dual to fluid dynamics does not require any such map for its perturbative construction and apparently there is no guarantee that the particular map used in `large-$D$' technique,  would also exist for the dynamical black-brane geometries, constructed in `derivative expansion'.\\
 In this note,  we have shown  that  the `hydrodynamic metric'\footnote{In this  note, the black-brane solution dual to fluid dynamics would always be referred to as the `hydrodynamic metric'.}   indeed could be `split' as required through an explicit computation upto first order in derivative expansion. This map could be constructed in any number of dimension and is independent of the `large -$D$' approximation. After determining this map, we have matched these two different gravity solutions upto the first subleading order on both sides. We believe  it would get more non-trivial at next order but we leave that for future.

One interesting outcome of this exercise, is the matching of the  dual theories of  both sides. It essentially reduces to a rewriting of hydrodynamics in large number of dimensions, in terms of the dynamics of the membrane. After implementing the correct gauge transformation, we finally get a field redefinition of the fluid variables (i.e., fluid  velocity and the temperature) in terms of membrane velocity and its shape\footnote{Truly speaking, what we have actually worked with is the reverse of what we have stated here, i.e., we determined the membrane velocity and the shape in terms of fluid variables, upto corrections of order ${\cal O}\left({1\over D},~\partial^2\right)$. This is just for convenience. The relations we found are easily invertible within perturbation. }.
We hope such a rewriting would lead to some new ways to view fluid and membrane dynamics and more ambitiously to a new duality between fluid and membrane dynamics in large number of dimensions, where gravity has no role to play (See \cite{AmosTurbulence}, \cite{Dandekar:2017aiv} for a similar discussion on such field redefinition and rewriting of fluid equations though in   \cite{AmosTurbulence} the authors have taken the large $D$ limit in a little different way than ours).

The organization of this note is as follows.\\
In section - (\ref{sec:overlap}) we first discussed about the overlap regime of these two perturbation schemes. Next in section - (\ref{sec:duality}) we discussed the map between the bulk  of the `black-hole' space-time and  the pure AdS, mentioned above and described an algorithm to construct the map, whenever it exists.  In section - (\ref{sec:hydromapex}) we compared the two metric and the two sets of dual equations (controlling the fluid-dynamics and the membrane dynamics respectively) within the overlap regime, upto the first subleading order on both sides. This section contains the main calculation of the paper. We worked out the map between these two sets of dual variables, leading to a map between large $D$ relativistic hydrodynamics and the membrane dynamics. Finally in section - (\ref{sec:conclude}) we concluded and discussed the future directions.

\section{The overlap regime}\label{sec:overlap}
In this section we shall discuss whether we could apply both `derivative expansion' and $\left({1\over D}\right)$ expansion simultaneously.
We shall first define the perturbation parameters for each of these two techniques in a precise way and also fix the range of their validity. We shall see that  these  two parameters  are completely independent  of each other and therefore their ratio could be tuned to any value, large or small.

Next we shall compare the  forms of the two metrics,  determined  using these two techniques, assuming the ratio (between the two perturbation parameters) to have any arbitrary value.

\subsection{Perturbation parameter in `derivative expansion'}\label{subsec:deriexpan}
Here we shall very briefly describe the method of `derivative expansion'. See \cite{Hubeny:2011hd} for a more elaborate discussion.\\

The technique of `derivative expansion' could be applied to construct a certain class of solutions to Einstein's equations in presence of negative cosmological constant in arbitrary dimension $D$.  
\begin{equation}\label{eineq}
\begin{split}
&~~~~\textit{The key gravity equation:}~~~\\
&{\cal E}_{AB} \equiv R_{AB} +(D-1) \lambda^2 g_{AB} =0\\
\end{split}
\end{equation}
$\lambda$ is the inverse of AdS radius. From now on, we shall choose  units such that $\lambda$ is set to one.\\
These gravity solutions are of  black hole' type, meaning they would necessarily have a singularity shielded by some horizon\cite{Bhattacharyya:2008xc}. They are in one-to-one correspondence with the solutions of relativistic Navier-Stokes equation in $(D-1)$ dimensional flat space-time (without any restriction on the value of $D$). In fact, we could use the hydrodynamic variables themselves to label the different gravity solutions, constructed using this technique of `derivative expansion'.
The labeling hydrodynamic variables are
\begin{enumerate}
\item Unit normalized velocity: $u^\mu (x)$ 
\item Local temperature: $T(x) = \left(D-1\over 4\pi\right) r_H(x)$
\end{enumerate}
At the moment $r_H$ is just some arbitrary length scale, which would eventually be related to the horizon scale of the dual black-brane metric.\\
$\{x^\mu\},~~~\mu = \{0,1,\cdots,D-2\}$ are the coordinates on the flat space-time whose metric is simply given by the Minkowski metric, $\eta_{\mu\nu} = Diag\{-1,1,1,1\cdots\}$.\\

`Derivative expansion' enters right into the definition of the  hydrodynamic limit. The velocity and the temperature of a fluid are functions of space-time but the functional dependence must be slow with respect to the length scale $r_H(x)$. For a generic fluid flow at a generic point, it implies the following.

Choose an arbitrary point $x^\mu_0$ ; scale the coordinates (or set the units) such that in the transformed coordinate $r_H(x_0) =1$. Now the technique of derivative expansion would be applicable  provided in this scaled  coordinate system 
\begin{equation}\label{validity}
\begin{split}
&\vert\tilde\partial_{\alpha_1}\tilde\partial_{\alpha_2}\cdots\tilde\partial_{\alpha_n} r_H\vert_{x_0}<<\vert\tilde\partial_{\alpha_1}\tilde\partial_{\alpha_2}\cdots\tilde\partial_{\alpha_{n-1}}r_H\vert_{x_0}<<\cdots<<\vert\tilde\partial_{\alpha_1}r_H\vert_{x_0}<<1~~~~\forall~{n,\alpha_i,x_0}\\
&\vert\tilde\partial_{\alpha_1}\tilde\partial_{\alpha_2}\cdots\tilde\partial_{\alpha_n} u^\mu\vert_{x_0}<<\vert\tilde\partial_{\alpha_1}\tilde\partial_{\alpha_2}\cdots\tilde\partial_{\alpha_{n-1}}u^\mu\vert_{x_0}<<\cdots<<\vert\tilde\partial_{\alpha_1}u^\mu\vert_{x_0}<<\vert u^\mu\vert~~~~\forall~{n,\alpha_i, x_0}
\end{split}
\end{equation}
In other words, the number of $\partial_\alpha$ derivatives  in a given term determines how suppressed the term is\footnote{The conditions as described in \eqref{validity} are for a generic situation. For a particular fluid profile, it could happen that at a given point in space-time  some $n$th order term is comparable to or even smaller than some $(n+1)$th order term. One might have to rearrange the fluid expansion around such anomalous points if they exist,  but they do not imply a `breakdown' of hydrodynamic approximation. As long as all derivatives in appropriate dimensionless coordinates are suppressed compared to one, `derivative expansion' could be applied.}.
In terms of original $x^\mu$ coordinates, each derivative $\partial_\mu$  corresponds to $r_H\tilde\partial_\mu$. Therefore  if we work in $x^\mu$ (which, unlike $\tilde x^\mu$, are not defined around any given point) coordinates, the parameter that controls the perturbation  is schematically $\sim r_H^{-1}\partial_\mu$.
\footnote{For a conformal fluid in finite dimension, there is only one length scale, set by the local temperature which also sets the scale of derivative expansion. But if we take $D\rightarrow\infty$, $T(x)$ and $r_H\sim {T(x)\over D}$ are two parametrically separated scales and  it becomes important to know  which one among these two scales controls the derivative expansion. In the condition \eqref{validity} we have chosen $r_H$ to be the relevant scale and set it to order ${\cal O}(1)$. 
Indeed  the results in  \cite{arbitrarydim} seem to indicate that terms of different derivative orders in hydrodynamic stress tensor, dual to gravity are  weighted by factors of $r_H\sim {T(x)\over D}$, and not $T$ alone.\\
Note that here the temperature of the fluid would scale as $D$, which is different from the  $D$ scaling of the temperature, imposed in \cite{AmosTurbulence}.}.

The starting point of this perturbation is a  boosted black-brane in asymptotically AdS space. The metric has the following form\\
 (in coordinates denoted as $\{r,x^\mu\} ,~~\mu =\{0,1,\cdots,D-2\}$. Units are chosen so that dimensionful constant, $\lambda$, appearing in equation \eqref{eineq} is set to one)\footnote{Note that the scaling of $\lambda$ with $D$ is upto us. At finite $D$ it is of no relevance, but it matters while taking the large $D$ limit. Here $\lambda$ would  be fixed to one as we would take $D$ to $\infty$.}.
\begin{equation}\label{eq:fluidstart2}
\begin{split}
& ds^2= - 2 u_\mu dx^\mu dr -r^2 ~f\left(r/r_H\right)~ u_\mu u_\nu dx^\mu dx^\nu + r^2 P_{\mu\nu} dx^\mu dx^\nu\\
&\text{where}~~~
f(z) = 1 - {1\over z^{D-1}},~~P_{\mu\nu} = \eta_{\mu\nu }+ u_\mu u_\nu
\end{split}
\end{equation}
Equation \eqref{eq:fluidstart2} is an exact solution to  equation \eqref{eineq} provided $u_\mu$ and $r_H$ are constants. \\
Now the algorithm for `derivative expansion' runs as follows. Suppose, $u^\mu$ and $r_H$ are not constants but are functions of $\{x^\mu\}$ .  Equation \eqref{eq:fluidstart2} will no longer be a solution. If we evaluate the gravity equation ${\cal E}_{AB}$ on \eqref{eq:fluidstart2}, the RHS will  certainly be proportional to the derivatives of  $u_\mu$ and $r_H$.  But $u_\mu$ and $r_H$ being the hydrodynamic variables, their derivatives are `small'  at every point in the sense described in \eqref{validity}. Therefore a `small' correction in the leading ansatz could solve the equation.\\
The $r$ dependence of these `small corrections ' could be determined exactly while the $\{x^\mu\}$ dependence would be treated in perturbation in terms of the labeling data $u^\mu(x)$ and $r_H(x)$ and their derivatives. $u^\mu(x)$ and $r_H(x)$ themselves would be constrained to satisfy the hydrodynamic equation, order by order in derivative expansion. While dealing with the full set of  gravity equations \eqref{eineq}, these equations on the hydrodynamic variables or the labeling data would emerge as the `constraint equations'  of the theory of classical gravity.

\subsection{Perturbation parameter in $\left(1\over D\right)$ expansion}\label{subsec:1/Dexp}
This is a perturbation technique, which is applicable only in a large number of space-time dimension (denoted as $D$), as a series expansion in powers of $\left(1\over D\right)$.  Clearly $\left(1\over D\right)$ is the perturbation parameter (a dimensionless number to begin with) here, which must satisfy 
$$\left(1\over D\right)<<1$$
Unlike the derivative expansion, the $\left(1\over D\right)$  expansion does not necessarily need the presence of cosmological constant, but we could also apply it if the cosmological constant is present provided we keep $\lambda$, the AdS radius (see equation \eqref{eineq} in subsection - \ref{subsec:deriexpan}) fixed as we take $D$ large. Note that the choice  $\lambda=1$, as we have done in previous subsection, is  consistent with this `$D$- scaling'.

The starting point here is the following metric.
\begin{equation}\label{eq:Dstart}
\begin{split}
&dS^2 \equiv{\cal G}_{AB}~ dX^A dX^B =\bar G_{AB} ~dX^A dX^B  + \psi^{-D} (O_A ~dX^A)^2\\
\end{split}
\end{equation}
where $\bar G_{AB}$, $\psi$ and $O_A$ are defined as follows.
\begin{enumerate}
\item $\bar G_{AB} $ is a smooth metric of pure AdS geometry which we shall refer to as `background'.\\
 We could choose any coordinate as along as the metric is smooth and all components of the Riemann curvature tensors are of order ${\cal O}(1)$ or smaller in terms of large $D$ - order counting.
\item  $\left(\psi^{-D}\right)$ is a harmonic function with respect to the metric $\bar G_{AB} $.  
\item $O_A$ is a null geodesic in the background satisfying $O_A n_B ~\bar G^{AB} =1$\\
where $n_A$ is the unit normal on the constant $\psi$ hypersurfaces  (viewed as hypersurfaces embedded in the background).
\end{enumerate}

The  metric \eqref{eq:Dstart} would solve the Einstein's equations \eqref{eineq} at leading order (which turns out to be of order ${\cal O}(D^2)$) provided the divergence of the ${\cal O}(1)$ vector field,  $U^A\equiv n^A-O^A$ with respect to the background metric is also of order ${\cal O}(1)$.
\begin{equation}\label{eq:leadconst}
\begin{split}
&\nabla\cdot U\equiv\bigg(\nabla\cdot n -\nabla\cdot O\bigg)_{\psi =1} = {\cal O}\left(1\right)\\
&\text{where $\nabla\equiv$  covariant derivative w.r.t. $\bar G_{AB}$}
\end{split}
\end{equation}
Naively  equation \eqref{eq:leadconst} does not seem to constrain the vector field $U^A$ since each of its components along with their derivatives in every direction are of order ${\cal O}(1)$ (this is what we mean by an `order ${\cal O}(1)$ vector field').  However, it is indeed a constraint  within the  validity-regime of  $\left(1\over D\right)$ expansion.  We could apply large $D$ techniques provided for a generic ${\cal O}(1)$ vector field $V^A\partial_A$, its divergence is of order ${\cal O}(D)$\footnote{This requirement certainly restricts the allowed dynamics that could be handled using this method.  But it is not as restrictive as it might seem to begin with. To see it explicitly, let us choose a coordinate system $\{z, y^\mu\}$ for the background.
\begin{equation}\label{eq:orderD}
\begin{split}
\bar G_{zz} =&~ {1\over z^2},~~~\bar G_{\mu\nu} = z^2\eta_{\mu\nu}\,~~~\text{Det}[\bar G]= -z^{(D-2)}\\
\nabla\cdot V=&~ z^{-(D-2)}\partial_z\left[z^{(D-2)} V^z\right] + \partial_\mu V^\mu\\
=&~\partial_z V^z + \partial_\mu V^\mu +(D-2)\left(V_z\over z\right) 
\end{split}
\end{equation}
Here clearly the first term is  of order ${\cal O}(1)$. The second term could potentially be of order ${\cal O}(D)$ since large number of indices are summed over. Still to precisely cancel against the last term, which certainly is of order  ${\cal O}(D)$ as long as $\left(V_z\over z\right) $ is not very small, it requires some fine tuning. Equation \eqref{eq:leadconst} says that $U^A\partial_A$ is such a fine-tuned vector field.}. \\
One easy way to ensure such scaling  would be to assume that the dynamics is confined within a finite number of dimensions and the rest of the geometry is protected by some large symmetry\cite{arbBack}. \\
 From now on, we shall assume such symmetry to be present in all the dynamics we discuss, including the dual  hydrodynamics, labeling the different geometries constructed in `derivative expansion'. For example, we shall assume that the divergence of the fluid velocity $u^\mu$, which we shall denote by $\Theta (\equiv \partial_\mu u^\mu)$, is always of order ${\cal O}(D)$,  whereas the velocity vector itself is of order ${\cal O}(1)$.

Now we shall  briefly describe some general features of this leading geometry in \eqref{eq:leadstart}. See \cite{arbBack} for a detailed discussion.\\
Firstly note that with the above conditions, the hypersurface  $\psi=1$  becomes  null  and we could identify this surface with the event horizon of the full space-time.\\
Also, if one is finitely away from the $\psi =1$ hypersurface, the factor $\psi^{-D}$ vanishes for large $D$ and the metric reduces to its asymptotic form $\bar G_{AB}$.\\
Next consider the region of thickness of the order of ${\cal O}\left(1\over D\right)$ around $\psi =1$ hypersurface.
This is the region\footnote{Following \cite{arbBack} , we shall refer to this region as `membrane region'}, where  $\left(1\over D\right)$  expansion would lead to a nontrivial correction to the leading geometry. To see why, let us do the following coordinate transformation.
$$X^A =X^A_0+ {\tilde x^A\over D}~~~\partial_A = D~\tilde\partial_A$$
where $\{X^A_0\}$ is an arbitrary point on the $\psi =1$ hypersurface.
In these new coordinates 
\begin{equation}\label{eq:Dstart2}
\begin{split}
&dS^2 =D^2G_{AB}~d\tilde x^A d\tilde x^B,~~~\text{where}~~~G_{AB}={\cal G}_{AB}\left(X_0 +{\tilde x\over D}\right)\\
\end{split}
\end{equation}
Now, if $\tilde x^A$ is not as large as $D$, it is possible to expand $\psi^{-D}$,  $O_A$ and $\bar G_{AB}$ around $X^A_0$. 
\begin{equation}\label{eq:1byd1}
\begin{split}
&\psi^{-D}(X^A) = e^{-\tilde x^AN_A}+ {\cal O}\left(1\over D\right),~~~\text{where}~~~N_A = \left[\partial_A \psi\right]_{X^A_0}\\
&O^A(X)= O^A\vert_{X^A_0} + + {\cal O}\left(1\over D\right),~~~G_{AB} (X) = G_{AB}\vert_{X^A_0} + {\cal O}\left(1\over D\right)
\end{split}
\end{equation}
Note that from the second condition (see the discussion below equation \eqref{eq:Dstart}) it follows that 
$$\text{Extrinsic curvature of $(\psi=1)$ surface} = K\vert_{\psi =1} = D\sqrt{N_A N_B \bar G^{AB}} + {\cal O}(1)$$
Substituting equation \eqref{eq:1byd1} in equation \eqref{eq:Dstart2} we find
\begin{equation}\label{eq:leadstart}
\begin{split}
G_{AB} =&~ O_A (X_0)~n_B(X_0)  +O_B(X_0)~n_A(X_0)+P_{AB} (X_0) \\
&-\left(1- e^{-\tilde x^AN_A} \right)~O_A(X_0) ~O_B(X_0)+ {\cal O}\left(1\over D\right)\\
&\text{where} ~~P_{AB}(X^0) \equiv \text{projector perpendicular to $n_A(X_0)$ and $O_A(X_0)$}\\
&n_A =\frac{\partial_A\psi}{\sqrt{(\partial_A\psi)(\partial_B\psi)\bar G^{AB}}}
\end{split}
\end{equation}
Clearly at the very leading order, the metric will have non-trivial variation only along the direction of $N_A$ - the  normal to the $\psi =1$ hypersurface at point $X^A_0$. Variations along all other directions are suppressed by factors of $\left(1\over D\right)$. This is very similar to the metric in equation \eqref{eq:fluidstart2} where at leading order the non-trivial variation is only along a single direction -  $r$. Therefore, within this `membrane region',  $\left(1\over D\right)$ expansion would {\it almost }reduce to derivative expansion  along directions other than $N_A$ provided the  metric \eqref{eq:leadstart} solves equation\eqref{eineq}  at very leading order. The conditions, listed below equation \eqref{eq:Dstart} along with equation \eqref{eq:leadconst}  ensure that this is the case.\\ Once the leading solution is found, the same algorithm, described in the previous subsection, would work and we could find the subleading corrections handling the variations of $N_A$ and $O_A$ along the constant  $\psi$ hypersurface. All such variations would be suppressed as long as  none of the components of $N_A$, $O_A$ and their derivatives (in the unscaled $X^A$ coordinates) are  as large as $D$. In other words, we should be able choose a coordinate system, along the horizon (or the hypersurface $\psi =1$) such that
\begin{equation}\label{eq:valid1}
\begin{split}
\left[{\bar G^{AB}\left(\partial_A~\psi^{-D}\right)\left(\partial_B~\psi^{-D}\right)}\right]^{-{1\over2}}\partial_A ~\vert_\text{horizon}<<1
\end{split}
\end{equation}
It is enough to impose this inequality only on the $\psi=1$ hypersurface;  the conditions listed below equation \eqref{eq:Dstart} will ensure that they are true  on all constant $\psi$ surfaces.\\

These conditions also specify the defining data (analogue of fluid-velocity  and temperature in case of  `derivative expansion') for the class of metrics, generated by $\left(1\over D\right)$  expansion. Here, the gravity solutions are expressed in terms of the auxiliary function $\psi$ and the one-form $O_A ~dX^A$. These two auxiliary fields satisfy the second and the third  conditions, listed below equation \eqref{eq:Dstart}. However, the above mentioned conditions, being differential equations, could  not fix the fields completely unless some boundary conditions are specified  along any fixed surface. The most natural choice for this  hypersurface is the surface given by $\psi=1$, which, by construction,  is the horizon of the full space-time geometry. Different metric solutions are classified by the shape of this surface and the components of $O_A$ projected along the surface.  Just as in `derivative expansion', we could solve for the metric correction only if these defining data (the projected $O_A$ field and the shape of the surface, encoded in its extrinsic curvature) satisfy the constraint equation, which we shall refer to as the `membrane equation'.

\subsection{Comparison between  two perturbation schemes}\label{subsec:compare}
In subsection-(\ref{subsec:1/Dexp}), we have seen that within the membrane region, ${\cal O}\left(1\over D\right)$ expansion is  {\it almost } like `derivative expansion' as described in subsection-(\ref{subsec:deriexpan}). Still it is also clear that they are not quite the same. The leading ansatz itself looks quite different for the two schemes, and there is no question of overlap if these two techniques compute perturbations around two entirely different geometries. So, to find an `overlap regime',  the first step would be to see  where in the parameter-space and in what sense, equation \eqref{eq:fluidstart2} and \eqref{eq:Dstart2} describe the same leading geometry.\\
Note that though the leading geometries look different algebraically, they both have  similar geometric properties -  namely the existence of a curvature singularity. In metric \eqref{eq:fluidstart2} it is located at $r=0$ and the metric \eqref{eq:Dstart2} is singular at $\psi =0$. Also the singularity is 
    shielded by some event-horizon\footnote{So far, the way both the techniques of `large-$D$ expansion' and `derivative expansion' are developed, the existence of a horizon is a must. It would be interesting to know whether we could depart from this condition and still apply either of these two techniques to construct `horizon free' or non-singular smooth geometries.}.\\
To see the similarities more explicitly, let us first choose a coordinate system  $X^A \equiv \{\rho,X^\mu\}$, such that the background metric- $\bar G_{AB}$  in equation \eqref{eq:1byd1} takes the form
\begin{equation}\label{eq:bspc}
\begin{split}
\bar G_{AB} ~dX^A~dX^B = {d\rho^2\over \rho^2} + \rho^2\eta_{\mu\nu} dX^\mu dX^\nu,~~~~~
\end{split}
\end{equation}
In this coordinate system, the following metric is an exact solution of equation \eqref{eineq}
\begin{equation}\label{eq:kerr}
\begin{split}
ds^2 = {d\rho^2\over \rho^2} + \rho^2\eta_{\mu\nu} dX^\mu dX^\nu + \left(\rho\over r_H\right)^{-(D-1)}\left( {d\rho\over \rho} - \rho~ dt\right)^2
\end{split}
\end{equation}
This is simply the Schwarsczchild black-brane solution, written in Kerr-Schild form.  Now let us note the following features of this metric\cite{arbBack}.
\begin{itemize}
\item The function $\left(\rho\over r_H\right)^{-(D-1)}$ is  harmonic with respect to the background upto correction of order ${\cal O}\left(1\over D\right)^2$.
$$\nabla^2\left(\rho\over r_H\right)^{-(D-1)} = {\cal O}\left(1\over D\right)^2$$
Hence the function $\left(\rho\over r_H\right)^{-(D-1)}$ could be identified with $\psi^{-D}$ appearing in the metric \eqref{eq:Dstart}    upto corrections of order ${\cal O}\left(1\over D\right)^2$.
\item The one form $\left( {d\rho\over \rho} - \rho~ dt\right)$ is null and satisfies the geodesic equation.
Further, contraction of this one-form with the unit normal to constant $\rho$ hypersurfaces is one.\\
Hence this one form could be identified with the null one form $O_A dX^A$
\end{itemize}
Hence it follows that the metric in \eqref{eq:kerr}, which is an exact solution of \eqref{eineq}, could be cast in the form of our leading ansatz upto  corrections subleading in $\left(1\over D\right)$ expansion. We could also expand the metric in equation \eqref{eq:kerr} around a given point on the horizon $\rho =r_H$, the same way we have done (see equation \eqref{eq:leadstart}) in the previous subsection with  the following set of identifications.
\begin{equation}\label{eq:identi}
\begin{split}
&N_A~dX^A\vert_{\rho =1} ={d\rho\over r_H},~~O_A~ dX^A\vert_{\rho =1} = {d\rho\over r_H} -r_H~dt\\
&n_A ~dX^A = {N_A~dX^A\over \sqrt{N_A N^A}}={d\rho\over r_H}\\
\end{split}
\end{equation}
The very leading term in this expansion, once written in terms of $N_A$ and $O_A$ would have exactly the same form as that of the metric  in equation \eqref{eq:Dstart2}.  The main difference between our leading ansatz,  equation \eqref{eq:Dstart} and equation \eqref{eq:kerr} is that in the later $N_A$ and $O_A$  satisfy equation \eqref{eq:identi} everywhere along the horizon, in the same $\{\rho, y^\mu\}$ coordinates.
For our leading ansatz  \eqref{eq:Dstart} also, it is true that  we could always choose a local $\{\rho, t\}$ coordinates by reversing the equations in \eqref{eq:identi}.  But for a generic $\psi$ and $O_A$,  this could not be done globally and this is the reason why our leading ansatz is not an exact solution of \eqref{eineq}.  However, the deviation  from the exact solution would clearly be proportional to the derivatives of $N_A$ and $O_A$ and therefore subleading.
So finally we conclude that  locally around a point on the horizon,  the leading ansatz for $\left(1\over D\right)$ expansion looks like a Schwarzschild black-brane written in a Kerr-Schild form with the local $\rho$ and $t$  coordinates, respectively oriented along the direction of the normal $N_A$ and the direction $O_A$ projected along the membrane $\psi = 1$.\\

Now let us come to the leading ansatz for the metric in derivative expansion. As it is explained in detail in \cite{nonlinfluid}, the leading ansatz in derivative expansion, equation \eqref{eq:fluidstart2} , reduces to Schwarzschild black-brane in Eddington-Finkelstein coordinates if we choose $r_H = constant$ and $u^\mu = \{1,0,0,\cdots\}$. Also  locally at any point $\{x^\mu_0\}$, we could always choose a coordinate system such that $u^\mu(x_0) = \{1,0,0,\cdots\}$, or in other words by appropriate choice of coordinates  locally the metric described in \eqref{eq:fluidstart2} could always be made to look like a Schwarzschild black-brane, though in a different gauge than in equation \eqref{eq:Dstart}.  Clearly the starting point of these different expansions are `locally 'same and it is possible to have an overlap regime.

But the difference lies in the concept of   `locality' and  also in the space of defining data.
In case of `large-$D$' expansion, the classifying data of the metric is specified on the horizon whereas for `derivative expansion' it is defined on the boundary of AdS space. \\
The range of validity for `large-$D$' expansion is given in equation \eqref{eq:valid1}. If we replace $\partial_A \psi^{-D}\vert_\text{horizon}$ by $ \left(-DN_A\right)$  the condition \eqref{eq:valid1} reduces to the existence of coordinate system such that
 \begin{equation}\label{eq:crit1byd}
\partial_A ~\vert_\text{horizon}<<D
 \end{equation}
which looks very similar to the  validity regime for `derivative expansion' , as already mentioned in subsection (\ref{subsec:deriexpan})
 \begin{equation}\label{eq:critderi}
 r_H^{-1}\partial_\mu<<1
 \end{equation}
If we could somehow map each point on the boundary to a point on the horizon (viewed as a hypersurface embedded in the background), the same $\{x^\mu\}$ coordinates could be used as coordinates along the horizon. In that case, whenever $r_H$ is of order ${\cal O}(1)$ in terms of `large-$D$'  order counting, the inequality \eqref{eq:critderi} would imply equation \eqref{eq:crit1byd}.
In other words, as $D\rightarrow \infty$,  all solutions of `derivative expansion' could be legitimately expanded further in $\left({1\over D}\right)$, though the reverse may not be true. 
 
Now we know that $\partial_A$ and $\partial_\mu$ are simply related (without any extra factor of $D$) for the case of exact Schwarzschild black-brane solutions. This is just the well-known coordinate transformation one should use to go from Kerr-Schild to Eddington-Finkelstein form of the black-brane metric. This transformation also gives  the required  map from the horizon to boundary coordinates. Once perturbations are introduced on both sides, we expect the  relation between  these two sets of coordinate systems would get corrected, but in a controlled and perturbative manner, thus maintaining the above argument for the existence of overlap.\\
 
So in summary, there does exist a region of overlap between these two perturbative techniques. In this note, our goal is to match them in the regime of overlap. As it is clear from the above discussion, the key step involves determining the map between $\partial_A$ and $\partial_\mu$, which we are going to elaborate in the next section.

\section{Transforming to `large-$D$' gauge}\label{sec:duality}

From the discussion of section - (\ref{sec:overlap}) it follows that if the space-time dimension $D$ is very large, we could always apply `$\left(1\over D\right)$ expansion' whenever `derivative expansion' is applicable.  Therefore a metric, corrected in derivative expansion in arbitrary dimension, when further expanded in $\left(1\over D\right)$, should reproduce the metric generated independently using the method of `$\left(1\over D\right)$ expansion'. More precisely if we take the metric of equation (4.1) from \cite{arbitrarydim} and expand it in $\left(1\over D\right)$,  it should match with the metric given in equation (8.1) of \cite{arbBack} after appropriate  transformation. \\
In this section our goal is to understand what these `appropriate  transformations' are.\\

Let us explain it in little more detail.\\
As we have mentioned before,  both of these  two perturbative techniques generate black brane geometries, in terms of a set of `dynamical data' , confined to a co dimension one hypersurface.  In the first case it is the boundary of the Asymptotic AdS space and in the second case it is the event horizon viewed as a hypersurface embedded in pure AdS. So both the techniques require a map from the full space-time geometry to a co dimension-one membrane.\\
The details of this map are quite clear for the case of `derivative expansion'.\\
The data-set that distinguishes between different dynamical geometries, here  is the profile of a relativistic conformal  fluid (its velocity and temperature). In other words, given a unit normalized velocity field and temperature, defined on a $(D-1)$ dimensional flat space-time and satisfying the relativistic Navier-Stokes equation, we should be able to uniquely construct  a $D$ dimensional space-time with a dynamical event horizon such that its metric is a solution to  \eqref{eineq}. The $(D-1)$ dimensional space is identified with the conformal boundary of this $D$ dimensional  black-brane geometry, which we shall refer to as bulk. This construction\cite{arbitrarydim} uses a very specific coordinate system, that encodes how a point in the bulk could be associated with a point in the boundary.
In \cite{Bhattacharyya:2008ji}, the authors have also explained how to reverse the construction of \cite{nonlinfluid},\cite{arbitrarydim}.  They have given an algorithm to read off the dual fluid variables starting from any black-brane geometry that admits derivative expansion, but  written in arbitray coordinates. This  explicitly proves the claim of one-to-one correspondence between the  dynamical black-brane geometry, admitting derivative expansion and the  fluid profile, satisfying relativistic Navier-Stokes equation. This algorithm has been heavily used to  cast the rotating black-holes in the `hydrodynamic form' \cite{arbitrarydim}.   \\

Similarly according to \cite{arbBack}, there exists a one-to-one correspondence between dynamical black-brane geometries in $\left(1\over D\right)$ expansion and a codimension-one `membrane dynamics'  in pure AdS space, though   \cite{arbBack} shows the correspondence in only one direction. It starts from a valid membrane data and integrate it outward towards infinity to construct the corresponding black-brane geometry. But to explicitly show this correspondence, we also need to know the reverse. In other words, we should know how to associate a point on the membrane to a point on the bulk and how to read off the membrane data,  starting from a dynamical black-brane geometry that admits an expansion in $\left(1\over D\right)$, but written in  some arbitrary coordinates.

In the  next subsection we  shall formulate an algorithm to determine this `membrane-bulk map',  analogous to the discussion of \cite{Bhattacharyya:2008ji} in the context of transforming the rotating black holes to the hydrodynamic form.

\subsection{Bulk-Membrane map}\label{subsec:mapmembrane}
\footnote{This subsection has been worked out by Shiraz Minwalla in a different context. We sincerely thank him for explaining it in detail to us. This `bulk-membrane' map is  the key concept  needed for the required `matching' of the two perturbative gravity solutions. }\\
The `large-D expansion' technique, as developed in \cite{arbBack}, would always generate the dynamical black-brane metric $G_{AB}$ in a `split' form. This `split' is specified in terms of an auxiliary function $\psi$ and an auxiliary vector field $O^A\partial_A$.  In terms of equation,
\begin{equation}\label{eq:schemeD}
\begin{split}
&G_{AB} = \bar G_{AB} +G_{AB}^{(\text{rest})}\\
\end{split}
\end{equation}
where $\bar G_{AB}$  is the background and $G_{AB}^{(\text{rest})}$  is such that  there exists a  null geodesic vector field $O^A\partial_A$ in the background, satisfying
\begin{equation}\label{eq:condo}
O^A~G_{AB} = O^A~\bar G_{AB}~~\Rightarrow~~O^A~G^\text{(rest)}_{AB} =0
\end{equation}
The normalization of this null geodesic vector is determined in terms of the function $\psi$, defined as follows.
\begin{enumerate}
\item  $\left(\psi^{-D}\right)$ is a harmonic function with respect to the metric $\bar G_{AB} $.  
\item $\psi=1$ hypersurface, when viewed as an embedded surface in full space-time,  becomes the dynamical event  horizon. This is  how the boundary condition on $\psi$ is specified.
\end{enumerate}
After fixing $\psi$, the normalization of $O^A$ is fixed  through the following condition.
$$O^A n_A  =1.$$
where $n_A$ is the unit normal on the constant $\psi$ hypersurfaces  (viewed as hypersurfaces embedded in the background).

The equations \eqref{eq:schemeD} and \eqref{eq:condo} together   specify a map between two entirely different geometries,  with metric $\bar G_{AB}$ and $G_{AB}$ respectively, both satisfying equation \eqref{eineq}. 
So if we want to recast an arbitrary dynamical black-brane metric, which admits $\left(1\over D\right)$ expansion, in the form as described in \eqref{eq:schemeD}, the first step would be to figure out this map or  the `split' of the  space-time  between `background' and the `rest',  so that the equation \eqref{eq:condo} is obeyed.\\

Now from the discussion of the previous subsection, we see that this `map' is crucially dependent on the vector field $O^A\partial_A$  and the function $\psi$. But both of them are defined using the `background' geometry and  we immediately face  a problem, since given an arbitrary black-brane metric,  it is the `background' that we are after.\\
 For example, given a black-brane metric we could always determine the location of the event horizon, but we would never know its embedding in the background, unless we know the `split' and therefore we would not be able to construct the $\psi$ function, by exploiting the harmonicity condition on $\psi^{-D}$. If we do not know $\psi$ we would not be able to orient or normalize  $O^A$, as required.\\ 

So we must have some equivalent formulation of  this `split' just in terms of the full space-time metric. The following observation allows us to do it. We could show that if $G_{AB}$ admits a split between $\bar G_{AB}$ and $G^{(\text{rest})}_{AB}$ satisfying $O^AG^{(\text{rest})}_{AB} =0$, then the vector   - $O^A\partial_A $ , which is a null geodesic with respect to $\bar G_{AB}$, is also a null geodesic with respect to $G_{AB}$. \\
 
 {\textbf {Proof}:}\\
 We know that 
 $$(O\cdot\nabla)O^A= \kappa ~O^A$$ 
 where $\nabla$ denotes the covariant derivative with respect to $\bar G_{AB}$ and $\kappa$ is the  proportionality factor. We would like to show that  
 $$(O\cdot\bar\nabla)O^A\propto O^A,~~\text{where $\bar\nabla$ is covariant derivative w.r.t. $G_{AB}$}$$
 Suppose $\bar\Gamma^A_{BC}$ denotes the Christoffel symbol corresponding to $\bar\nabla_A$ and $\Gamma^A_{BC}$ denotes the Christoffel symbol corresponding to $\nabla_A$. These two would be related as follows \cite{arbBack}.
 \begin{equation}\label{eq:gamma}
 \begin{split}
 \bar\Gamma^A_{BC} = \Gamma^A_{BC} +\underbrace{ {1\over 2}\left(\nabla_B \left[ G^{(\text{rest})}\right]^A_C + \nabla_C\left[G^{(\text{rest})}\right]^A_B -\nabla^A \left[G^{(\text{rest})}\right]_{BC}\right)}_{\delta\Gamma^A_{BC}}
 \end{split}
 \end{equation}
 Here all raising and lowering of indices have been done using $\bar G_{AB}$.
Note that 
\begin{equation}\label{eq:simp1}
\begin{split}
O^BO^C~\delta\Gamma^A_{BC} = ~&O^B(O\cdot\nabla)\left[G^{(\text{rest})}\right]^A_B -{1\over 2} O^B O^C\nabla^A\left[G^{(\text{rest})}\right]_{BC} \\
=~&- \left[G^{(\text{rest})}\right]^A_B \left[ (O\cdot\nabla)O^B\right]  + {1\over 2} \left(\nabla^AO^C\right)\left[G^{(\text{rest})}\right]_{BC} O^B\\
=~&\kappa\left(O^C\left[G^{(\text{rest})}\right]^A_C\right) =0
\end{split}
\end{equation}
What we want to show simply follows from equation \eqref{eq:simp1}
\begin{equation}\label{proof1}
\begin{split}
(O\cdot\bar\nabla) O^A = (O\cdot\nabla) O^A = \kappa~O^A
\end{split}
\end{equation}

So we could determine $O^A$ by solving the null geodesic equation with respect to the full space-time metric $G_{AB}$. But to determine it fully, we also need to know $\kappa$, fixed by the normalization of $O^A$. As mentioned before, the normalization used previously in the application of `large-$D$' technique is not suitable for our purpose, since it requires the knowledge of the `background' beforehand.
 But luckily the  form of the `split', which is defined by the condition $\bigg[O^AG^{(\text{rest})}_{AB} =0\bigg]$ is independent of the normalization of $O^A$.\\
  So we shall first determine another null geodesic field (let us denote it by $\bar O_A$ to remind ourselves of  the difference in normalization) which is affinely parametrized and whose inner-product with the normal to event horizon (which, upto normalization,  could again be determined without any knowledge of the `split') is one.\\
   Now we are at a stage to define the map the  between the `background' and the full space-time geometry.

Suppose $\{Y^A\}$ denote the coordinates in the background geometry (in our case pure AdS, the metric is denoted by $\bar{G}_{AB}$) and $\{X^A\}$ are the coordinates of the full space-time (the dynamical black-brane,  the metric is denoted by ${\cal G}_{AB}$). Let us denote the invertible functions that give a one to one correspondence between these two spaces as $\{f^A\}$.\\
\begin{equation}\label{eq:map}
\begin{split}
&Y^A = f^A(\{X\})\\
\end{split}
\end{equation}
The equations that will determine $f^A$ s are the following
\begin{equation}\label{eq:mapdef}
\begin{split}
\bar O^A ~ {\cal G}_{AB}\vert_{\{X\}}= \bar O^A\left( {\partial f^C\over \partial X^A}\right)\left( {\partial f^{C'}\over \partial X^B}\right)\bar{G}_{CC'}\vert_{\{X\}}
\end{split}
\end{equation}
\footnote{The subscript $\{X\}$  in equation \eqref{eq:mapdef} denotes that both LHS and RHS of equation \eqref{eq:mapdef} have to be evaluated in terms $\{X^A\}$ coordinates.}
Here $\bar O^A$ are affinely parametrized the null geodesics in the full space-time geometries i.e.,
\begin{equation}\label{eq:barO}
 \bar O\cdot \bar\nabla \bar O^A =0
 \end{equation}
 Equation \eqref{eq:barO} would fix $\bar O_A$ completely once we specify  the angles it would make with the tangents of the horizons, which is effectively a set of $(D-1)$ numbers. Now what  we are actually interested in is not $\bar O_A$ but $O_A$ which is related to $\bar O_A$ with a normalization. Therefore we are free to choose the normalization of $\bar O_A$, since anyway we have to re-normalize it again. This will fix one of the $(D-1)$ initial conditions. Rest we shall keep arbitrary. \\
 We shall assume
\begin{equation}\label{eq:recond}
\begin{split}
&\bar O^A N_A\vert_\text{horizon} =1\\
&\bar O^A l^{(i)}_A\vert_\text{horizon} = \text{ some arbitrary functions of horizon cordinates}
\end{split}
\end{equation} 
 where $N_A $ is the null normal to the event horizon (with some arbitrary normalization)
  and $l_{(i)}^A\partial_A$ s are the unit normalized space-like tangent vectors to the horizon.\\
It turns out that the hydrodynamic metric could be split for a very specific choice of these spatial initial conditions and we shall fix them order by order in derivative expansion by matching the hydrodynamic and the `large-$D$' metric. 
Once $\bar O^A$  is fixed (in terms of these arbitrary angles),  we could determine $f^A$ s upto some integration constants by solving equation \eqref{eq:mapdef}.

 Equation \eqref{eq:mapdef} further says that if we apply the map \eqref{eq:map} as a coordinate transformation on the `background', then in the new $\{X^A\}$ coordinates the map would just be an `identity' map and the full space-time metric  ${\cal G}_{AB}$ would admit the split as given in equation \eqref{eq:schemeD} satisfying \eqref{eq:condo} \footnote{We would  also like to emphasize that what we are describing here is not just a gauge or coordinate  transformation. The `split' mentioned in equation \eqref{eq:schemeD} is a genuine point-wise map between two entirely different geometries. Once we have figured out the `map', we are free to transform the coordinates further; both $G_{AB}$ and $\bar G_{AB}$ would change, but the `map' will still be there.}.\\
 Once we have figured out how to split the full space-time metric into `background' and the `rest', we know how to view the event horizon as a surface embedded in the `background' and therefore the auxiliary function $\psi$ (by solving the harmonicity of $\psi^{-D}$ w.r.t the background) everywhere. Now we can normalize $\bar O^A$ as it has been done in \cite{arbBack}. Using these $\psi$ and $O^A$ (appropriately normalized) one should be able to recast any  arbitrary metric, that admits large-$D$ expansion, exactly in the form  of \cite{arbBack}.

\section{Bulk-Membrane map in metric dual to Hydrodynamics}\label{sec:hydromapex}
In this subsection, we shall implement the above algorithm, described in the previous subsection,  for the metric dual to hydrodynamics. For convenience we are summarizing the steps again.
\begin{itemize}
\item Determine the equation for the event horizon.
\item Determine the null normal to the horizon.
\item Solve equation \eqref{eq:barO}  to determine $\bar O^A$  everywhere. We need the normal, derived in previous step, to impose the boundary condition.
\item Choose any arbitrary coordinate system $\{Y^A\}$,  where the `background' has a smooth metric $G_{AB}$.
\item Now solve the equation \eqref{eq:mapdef} to determine the mapping functions $f^A$ 's.
\end{itemize}
 For a generic dynamical metric, it is not easy to implement all these steps. But in this case what would help us is the `derivative expansion' and the fact that $f^A$ 's are exactly known at zero derivative order; it is simply the coordinate transformation between Eddington-Finkelstein and Kerr-Schild form of a static black brane metric.
 
 Though the zeroth order transformation is already known, as a `warm-up' exercise we shall re-derive it using the above algorithm. The condition of `staticity' and translational symmetry of the metric allow us to solve relevant equations exactly in this case.
 
 \subsection{Zeroth order in `derivative expansion':}
At zeroth order in derivative expansion the metric dual to hydrodynamics has the following form
\begin{equation}\label{eq:zeroderi}
\begin{split}
& ds^{2}=-2 u_{\mu}{dx}^{\mu}dr -r^2f\left({r/r_H}\right)u_{\mu}u_{\nu}dx^{\mu}dx^{\nu}+r^{2}P_{\mu \nu}dx^\mu dx^\nu\\
&\text{where}~~~P_{\mu\nu} \equiv \eta_{\mu\nu}+u_\mu u_\nu,~~f(z) \equiv \left[1- z^{-(D-1)}\right],~~u_\mu u_\nu \eta^{\mu\nu} =-1
 \end{split}
\end{equation}
We could read off the components of the metric and its inverse.
\begin{equation}\label{eq:compmetinv}
\begin{split}
&{\cal G}_{rr}=0,~~~{\cal G}_{\mu r} = -u_\mu,~~~{\cal G}_{\mu\nu} =-r^2f\left({r/ r_H}\right)u_{\mu}u_{\nu}+r^{2}P_{\mu \nu}\\
&{\cal G}^{rr}=r^2f\left({r/ r_H}\right),~~~{\cal G}^{\mu r}=u^{\mu},~~~{\cal G}^{\mu\nu}=\frac{1}{r^{2}}P^{\mu\nu}\\
\end{split}
\end{equation}
At zero derivative order both $r_H$ and $u^\mu$ could be treated as constants,
The event horizon and the null normal to it are  given by
\begin{equation}\label{eq:zeroevent}
\text{Event Horizon}: {\cal S} = r -r_H =0,~~~N_A ~dX^A = dX^A \partial_A {\cal S}=dr
\end{equation}
Now we shall  figure out the `map' that will lead to the desired `split' between `background' and `rest'. 

We have already determined the event horizon.
Next we have to  solve for $\bar O^A$, satisfying the conditions
$$\bar O^B\bar\nabla_B\bar O^A =0,~~\bar O^A\bar O^B{\cal G}_{AB} =0,~~\bar O^AN_A\vert_{r=r_H}=\bar O^r\vert_{r=r_H}=1$$

At zero derivative order, ${\cal G}_{AB}$ has translational symmetry in all the $x^\mu$. The conditions on $\bar O^A$ does not break this symmetry. 
Hence  $\bar O^A$ must have the form
\begin{equation}\label{eq:Oansatz}
\bar O^A\partial_A = h_1(r)~ \partial_r + h_2(r)~ u^\mu\partial_\mu 
\end{equation}
Now we shall process the condition that  $O^A $ is a null vector field.
\begin{equation}\label{eq:Oprocess1}
\begin{split}
~ &\bar O^{A}\bar O^{B}{\cal G}_{AB}=0\\
 \Rightarrow~&2h_{2}(r)h_{1}(r){\cal G}_{\mu r}u^{\mu}+{h_{2}(r)}^{2}u^{\mu}u^{\nu}{\cal G}_{\mu\nu}=0\\
 \Rightarrow~&h_{2}(r)\left[2h_{1}(r)-r^2f\left({r/r_H}\right)h_{2}(r)\right]=0\\
 \Rightarrow~&h_{2}(r)=0
 \end{split}
\end{equation}
So finally $\bar O^A\partial_A = h_1(r)\partial_r$\footnote{Actually there are two solution to \eqref{eq:Oprocess1}. If we assume $h_{2}(r)\neq0$ and finite everywhere, then 
$$h_1(r) = ~{r^2\over 2}f\left({r / r_H}\right)h_{2}(r)$$
This implies that $h_1(r)$ will vanish at the horizon $r=r_H$ (which is a zero of the function $f\left({r/ r_H}\right)$), contradicting the boundary condition on $\bar O^r$.}.\\
Substituting this form of $\bar O^A$ in the geodesic equation we could see that $h_1(r)$ has to be a constant and then boundary condition simply says that $h_1(r) =1$
\begin{equation}
 \bar O^{A}\partial_{A}=\bar O^{r}\partial_{r}=\partial_{r}
\end{equation}

Now let us choose a coordinate system $Y^A=\{\rho,y^\mu\}$ for the `background'  where the metric takes the following form
\begin{equation}\label{eq:explicitbck}
 {ds}^{2}_{background}=\frac{{d\rho}^{2}}{{\rho}^{2}}+{\rho}^{2}\eta_{\mu\nu} ~dy^\mu ~dy^\nu
\end{equation}
Again the symmetries  motivate us to take the following form for the mapping, which gives the one to one correspondence between the  background coordinates $\{Y^A\} =\{\rho,y^\mu\} $ and black-brane coordinates $\{X^A\} =\{r,x^\mu\}$
\begin{equation}\label{eq:zeromap}
 y^{\mu}=x^{\mu}+g(r)u^{\mu},~~~
 \rho=h(r)
\end{equation}
Let us apply the map \eqref{eq:zeromap} as a coordinate transformation on the background. In the new coordinates (where the map is just an `identity') the background metric takes the following form
\begin{equation}\label{eq:zeronewcoord}
\begin{split}
\bar {\cal G}_{rr}= \left({h'\over h}\right)^2 - \left( g'h\right)^2,~~\bar{\cal G}_{\mu r} = g'h^2  u_\mu,~~\bar{\cal G}_{\mu \nu} = h^2~\eta_{\mu\nu} 
\end{split}
\end{equation}
Here we have suppressed the $r$ dependence and derivative w.r.t $r$ is denoted by prime ($'$).
In this coordinates equation \eqref{eq:mapdef} takes the form
\begin{equation}\label{eq:zeromapdef}
\begin{split}
\left({h'\over h}\right)^2 - \left( g'h\right)^2=0,~~g'h^2 =-1
\end{split}
\end{equation}
These two equation could be solved very simply. The general solution
\begin{equation}\label{eq:zeromapsol}
\begin{split}
h(r) = \pm (r + c_1),~~~g(r) = {1\over r+ c_1} +c_2
\end{split}
\end{equation}
where $c_1$ and $c_2$ are two arbitrary constants.\\
We shall choose the plus sign in $h(r)$ to make sure that whenever $r$ increases,  $\rho$ also  increases.

 Now we have to fix the integration constants.
Note that once we know the map, we know the form of ${\cal G}^\text{(rest)}_{AB}$, satisfying equation \eqref{eq:condo} by construction.
\begin{equation}\label{eq:zerorest}
\begin{split}
&{\cal  G}^\text{(rest)}_{rr}={\cal  G}^\text{(rest)}_{r\mu}=0 \\
&{\cal  G}^\text{(rest)}_{\mu\nu}=\left[(r+c_1)^2 - r^2f(r/r_H)\right]u_\mu u_\nu + \left[r^2- (r+c_1)^2\right]P_{\mu\nu}\\
\end{split}
\end{equation}
Now  we further want that if $D\rightarrow\infty$, the metric should reduce to its asymptotic form at any finite distance from the event  horizon or in other words,  ${\cal  G}^\text{(rest)}_{\mu\nu}$  must vanish outside the `membrane region' (a region with `thickness' of the order of ${\cal O}\left(1\over D\right)$ around the `membrane', see section (\ref{subsec:1/Dexp})). This condition will force us to set $c_1=0$. The other constant $c_2$ is not appearing in the final form of the metric at all, so this ambiguity will remain here at this order and it is simply a consequence of the translational symmetry  in $x^\mu$ and $y^\mu$ directions. For simplicity we shall also choose $c_2 =0$. So the final form of the map at zeroth order
\begin{equation}\label{zeromapfinal}
\rho = r,~~~~~y^\mu = x^\mu + {u^\mu\over r}
\end{equation}

\subsection{First order in derivative expansion}\label{subsec:1stderi}
In this subsection we shall extend the computation of the previous subsection upto the first order in derivative expansion.  Here $u^\mu$ and $r_H$ depends on $x^\mu$ but any term that has more than one derivatives of $u^\mu$ and $r_H$ has been neglected. All calculations presented in this subsection generically will have corrections at order ${\cal O}(\partial^2)$.

At first order in derivative expansion the metric dual to hydrodynamics has the following form \cite{arbitrarydim}
\begin{equation}\label{eq:firstderi}
\begin{split}
ds^{2}=&~~-2u_{\mu}dx^{\mu}dr-r^{2}f\left(r/ r_H\right)u_{\mu}u_{\nu}dx^{\mu}dx^{\nu}+r^{2}P_{\mu\nu}dx^{\mu}dx^{\nu}\\
&~~+r\bigg[-(u_{\mu}a_{\nu}+u_{\nu}a_{\mu})+\frac{2\Theta}{D-2} u_{\mu}u_{\nu}+2F\left(r/r_H\right)
{\sigma}_{\mu\nu}\bigg]dx^{\mu}dx^{\nu}\\
\end{split}
\end{equation}
Where, 
\begin{equation}\label{eq:integral}
 F(r)=r\int_{r}^{\infty}dx\frac{x^{D-2}-1}{x(x^{D-1}-1)}\nonumber
\end{equation}
And\footnote{Here `$\cdot$' denotes  contraction with respect to $\eta_{\mu\nu}$}
\begin{equation}
\begin{split}\label{eq:sigmadef}
a_\mu = (u\cdot\partial)u_\mu ~,~~~~\Theta = \partial\cdot u~,~~~~
 \sigma^{\mu\nu}=P^{\mu\alpha}P^{\nu\beta}\left(\partial_{\alpha}u_{\beta}+\partial_\beta u_\alpha\over 2\right) -\left(\frac{\Theta}{D-2}\right)P^{\mu\nu}
 \end{split}
\end{equation}
We shall often refer to this metric, described in equation \eqref{eq:firstderi}, as `hydrodynamic metric'. Here both $r_H$ and $u_\mu$ are functions of $x^\mu $s; but they are not completely arbitrary. the hydrodynamic metric will solve the Einstein's equations (upto corrections of order ${\cal O}(\partial^2)$) provided the derivatives of $r_H$ and $u_\mu$ satisfies the following equations\footnote{These two equations are just the stress tensor conservation equation  for a $(D-1)$ dimensional ideal conformal fluid. }.
\begin{equation}\label{eq:conservationeq}
\begin{split}
{(u\cdot\partial) r_H\over r_H} + {\Theta\over D-2}=0,~~~P^{\mu\nu}\left(\frac{\partial_\mu r_H}{r_H}\right) + a^\nu=0
\end{split}
\end{equation}
We read off the components of the metric and its inverse
\begin{equation}\label{eq:compmet}
\begin{split}
{\cal G}_{\mu r}=&~-u_{\mu}, ~~{\cal G}_{rr}=0\\
{\cal G}_{\mu\nu}=&~-r^{2}f\left(r/ r_H\right)u_{\mu}u_{\nu}+r^{2}P_{\mu\nu}\\
&~+r\bigg[-(u_{\mu}a_{\nu}+u_{\nu}a_{\mu})+\left(\frac{2\Theta}{D-2} \right)u_{\mu}u_{\nu}+2F\left(r/r_H\right)
{\sigma}_{\mu\nu}\bigg]
\end{split}
\end{equation}

\begin{equation}\label{eq:compinv}
\begin{split}
&{\cal G}^{rr} = r^2 f(r/r_H)-r\left(\frac{2\Theta}{D-2}\right) ,~~~~{\cal G}^{\mu r} = u^\mu -{a^\mu\over r}\\
&{\cal G}^{\mu\nu}=\frac{P^{\mu\nu}}{r^2}-\frac{2F(r/r_H)}{r^3} ~\sigma^{\mu\nu}
\end{split}
\end{equation}\\
The horizon is still given by the surface (no correction at first order in derivative, though the normal gets corrected since $\partial_\mu r_H$ is not negligible now.)
\begin{equation}\label{eq:zoneevent}
\text{Event Horizon}: {\cal S} = r -r_H =0,~~~N_A ~dX^A = dX^A \partial_A {\cal S}=dr - dx^\mu~\partial_\mu r_H
\end{equation}
We need the Christoffel symbols to compute the geodesic equation.
\begin{equation}\label{eq:christ}
\begin{split}
 &\Gamma_{rr}^{r}=0,~~~~~~\Gamma_{rr}^{\mu}=0\\
 &\Gamma_{\alpha r}^{r}=\left[rf(r/r_H)+\frac{r^2}{2r_H}f'(r/r_H)-\frac{\Theta}{D-2}\right]u_{{\alpha}}\\
 &\Gamma_{r\delta}^{\mu}=\frac{1}{2r^2}\left[2rP^{\mu}_{\delta}-\partial_{\delta}u^{\mu}-u_{\delta}a^{\mu}+\partial^{\mu}u_{\delta}+u^{\mu}a_{\delta}-2F(r/r_H)\sigma^{\mu}_{\delta}+2\left(r/r_H\right)F'(r/r_H)\sigma^{\mu}_{\delta}\right]\\
 \end{split}
\end{equation}
At first order in derivative expansion, the most general correction that could be added to  $\bar O^A$,  maintaining it as a null vector with respect to the first order corrected metric:
\begin{equation}
 \bar O^{A}\partial_{A}=\partial_{r}+w_{1}(r)~\Theta~\partial_{r}+w_{2}(r)~a^{\mu}\partial_{\mu}
\end{equation}
We shall fix $w_1(r)$ and $w_2(r)$  using the geodesic equation.\\
The $r$ component of the geodesic equation gives the following.
\begin{align*}
 &(\bar O\cdot\bar \nabla)\bar O^{r}=0\\
 \Rightarrow&\bar O^{r}\bar\nabla_{r}\bar O^{r}+\bar O^{\mu}\bar \nabla_{\mu}O^{r}=0\\
 \Rightarrow&\bar O^{r}\partial_{r}\bar O^{r}+\Gamma_{rr}^{r}\bar O^{r}\bar O^{r}+2\bar O^{r}\bar O^{\alpha}\Gamma_{\alpha r}^{r}=0\\
 \Rightarrow&(1+w_{1}(r)\Theta)w_{1}'(r)\Theta+2(1+w_{1}(r)\Theta)(w_{2}(r)a^{\alpha})\Gamma_{\alpha r}^{r}=0\\
 \Rightarrow&w_{1}'(r)=0\\
 \Rightarrow&w_{1}(r)=A_{1},~~~~\text{where $A_{1}$ is a constant}\\
\end{align*}

From  the $\mu$ component of the geodesic equation we find
\begin{align*}
&(\bar O\cdot\bar\nabla)\bar O^{\mu}=0\\
\Rightarrow&~\bar O^{r}
\bar \nabla_{r}\bar O^{\mu}+\bar O^{\lambda}\bar \nabla_{\lambda}\bar O^{\mu}=0\\
\Rightarrow&~\bar O^{r}\partial_{r}\bar O^{\mu}+\bar O^{r}\bar O^{r}\Gamma_{rr}^{\mu}+2\bar O^{r}\bar O^{\delta}\Gamma_{r\delta}^{\mu}=0\\
\Rightarrow&~\left[w_2'(r) +\frac{2w_2(r)}{r}\right]a^{\mu}=0\\
\Rightarrow&~w_2(r)=\left(\frac{A_{2}}{r^{2}}\right),~~~\text{where $A_2$ is another integration constant}
\end{align*}
At this stage
\begin{equation}
 \bar O^{A}\partial_{A}=\partial_{r}+A_{1}\Theta~\partial_{r}+\left(\frac{A_{2}}{r^{2}}\right)~a^{\mu}\partial_{\mu}
\end{equation}
We could partially fix the integration constants using the boundary conditions.\\
At horizon
\begin{equation}\label{eq:implbc}
\begin{split}
 &\bar O^{A}N_A\arrowvert_{r=r_H}=1~~ \Rightarrow~(1+A_{1}\Theta)=1\Rightarrow ~~A_1=0\\
 &\bar O^{\mu}\partial_{\mu}r_H={\cal O}\left(\partial^2\right)~\Rightarrow~\text{ No constraint on $A_2$}
 \end{split}
\end{equation}\\
Hence it follows that .
\begin{equation}\label{eq:finOa}
\begin{split}
&\bar O^{A}\partial_{A}=\partial_{r} +\left(\frac{A_{2}}{r^{2}}\right)~a^{\mu}\partial_{\mu}+ \text{terms 2nd order in derivative expansion}\\
\Rightarrow~&\bar O_A~ dX^A = -u_\mu ~dx^\mu +A_2~ a_\mu~dx^\mu+ \text{terms 2nd order in derivative expansion}\\
\end{split}
\end{equation}
Next we have to solve for the `mapping functions'.  Let us choose the same coordinates $\{Y^A\}$, as  in the previous subsection so that  background  takes the form  of  equation \eqref{eq:explicitbck}. We expect  that  the mapping functions  \eqref{zeromapfinal}  will get corrected by first order terms in derivative expansion.
\begin{align}\label{eq:1stmap}
 &y^{\mu}=x^{\mu}+\frac{u^{\mu}(x)}{r}+f_{1}(r)\Theta ~u^{\mu}(x)+f_{2}(r)~a^{\mu}(x),~~~
 \rho=r+f_{3}(r)~\Theta
\end{align}
As before,  we shall apply the map \eqref{eq:1stmap} as a coordinate transformation on the background. In the new coordinates (where the map is just an `identity') the background metric takes the following form
\begin{equation}\label{eq:1stnewcoord}
\begin{split}
&\bar {\cal G}_{rr}=2\left(f_1'(r) + { f_3'(r)\over r^2} -{2f_3(r)\over r^3}\right)~\Theta\\
&\bar{\cal G}_{\mu r} = -\left[1- \left(r^2f'_1(r)- {2f_3(r)\over r}\right)\Theta\right] u_\mu + r^2f_2'(r)~a_\mu\\
&\bar{\cal G}_{\mu \nu} =r^2 \left(1+{2  f_3(r)\over r}~\Theta\right) ~\eta_{\mu\nu}  + r\left(\partial_\nu u_\mu +\partial_\mu u_\nu\right)
\end{split}
\end{equation}

Substituting equation \eqref{eq:1stnewcoord} in equation  \eqref{eq:mapdef} we find
\begin{equation}\label{eq:1stmapdef}
\begin{split}
&\bar{\cal G}_{\mu r}  + \left(A_2\over r^2\right)a^\nu\bar{\cal G}_{\nu\mu}= -u_\mu + A_2~a_\mu + {\cal O}\left(\partial^2\right),~~~\bar{\cal G}_{rr}=0\\
\Rightarrow~&r^2f_1'(r) -{2f_3(r)\over r}=0,~~~f_2'(r) =0,~~~f_1'(r) + { f_3'(r)\over r^2} -{f_3(r)\over r^3}=0\\
\end{split}
\end{equation}
The general solution for equation \eqref{eq:1stmapdef}:
\begin{equation}\label{eq:1stsol}
\begin{split}
&f_3(r) = C_3,~~~~f_2(r) = C_2,~~~~f_1(r) = C_1 -{C_3\over r^2}\\
&\text{where $C_1$, $C_2$ and $C_3$ are arbitray constants}
\end{split}
\end{equation}
In the new $X^A = \{r,x^\mu\}$ coordinates the metric of the background takes the following form
\begin{equation}\label{eq:metback}
\begin{split}
ds_\text{background}^2
=~&\bar{\cal G}_{AB} dX^A dX^B \\
=& ~-2u_\mu dx^\mu~dr + r^2\eta_{\mu\nu} dx^\mu~dx^\nu \\
&~+r\left[2C_3\Theta~\eta_{\mu\nu} + (\partial_\mu u_\nu +\partial_\nu u_\mu)\right] dx^\mu dx^\nu\\
\\
=& ~-2u_\mu dx^\mu~dr + r^2\eta_{\mu\nu} dx^\mu~dx^\nu \\
&~+2r\left[-C_3\Theta ~u_\mu u_\nu +\left(C_3+{1\over D-2}\right)\Theta~P_{\mu\nu} - \left(a_\mu u_\nu +a_\nu u_\mu\over 2\right) + \sigma_{\mu\nu}\right] dx^\mu dx^\nu\\
\end{split}
\end{equation}
In the last step we have rewritten ${\cal G}_{\mu\nu}$ using the following identity
\begin{equation}\label{eq:idnt}
\begin{split}
\partial_\mu u_\nu +\partial_\nu u_\mu = 2\sigma_{\mu\nu} + \left(2\Theta\over D-2\right) P_{\mu\nu} - (a_\mu u_\nu + a_\nu u_\mu)
\end{split}
\end{equation}
Once we know the background, we could determine $\bar{\cal G}_{AB}^\text{rest}$.
\begin{equation}\label{eq:1strest}
\begin{split}
&{\cal G}^\text{(rest)}_{rr} =0,~~~{\cal G}^\text{(rest)}_{\mu r} =0\\
&{\cal G}^\text{(rest)}_{\mu \nu}= r^2\left(r_H\over r\right)^{D-1} u_\mu u_\nu-2r~\tilde C_3~\Theta~\eta_{\mu\nu}  +2r\left[F(r/r_H) -1\right]\sigma_{\mu\nu}\\
&\text{where}~~~\tilde C_3\equiv C_3 + {1\over D-2}
\end{split}
\end{equation}

\section{Hydrodynamic metric in $\left(1\over D\right) $ expansion}
In this section we would like to expand the `hydrodynamic metric'  (already split into `background' and `rest' in the previous secion) in an expansion in $\left(1\over D\right)$ and compare it against the metric  described in \cite{arbBack}. 

This comparison involves two steps. The first one is of course an exact match  of the  two metric upto the required order.
The second step involves the mapping of the evolution of  the data. Let us explain it in a little more detail.
\newline
As  we have mentioned before, both `hydrodynamic metric' and `large - $D$' metric are determined in terms of data, defined on a co dimension one hypersurfaces - in the first case it is the velocity and temperature of a relativistic fluid living on the boundary of asymptotic AdS and in the second case it is the horizon viewed as a membrane embedded in the background with fluctuating shape and velocity.
However we cannot choose the data arbitrarily. The hydrodynamic metric or the large $D$ metric will solve the Einstein's equations only if the corresponding data satisfy certain evolution equation. For matching of these two metrics, the evolution of the data also should match. More precisely , we should be able to re express the membrane velocity and shape in terms of fluid velocity and temperature and further we have to show that once hydrodynamic equations are satisfied, the membrane equation is also true upto the required order.

Below we shall first compare the two metrics and in the next subsection we shall prove the equivalence of the evolution of these two sets of defining data.

\subsection{ Comparison between the two metrics}
If the hydrodynamic metric has to match with the final metric described in \cite{arbBack}, the first requirement is that $\bar{\cal G}_{\mu\nu}^\text{rest}$ must vanish as one goes finitely away from the horizon. This is possible provided $\tilde C_3$ is  zero and also  the function $\left[F(r/r_H)-1\right]$  has a certain type of fall-off behavior   at large $r$. Now $\tilde C_3$ being an integration constant we could easily set it to zero.
In appendix (\ref{app:integral}) we have analyzed the integral \eqref{eq:integral} and therefore the function $\left[F(r/r_H)-1\right]$. It turns out that at large $D$ this integral could be approximated as follows.
\begin{equation}\label{eq:Fexprep}
\begin{split}
F(z) = F\left(1 +{Z\over D}\right) = 1 - \left(1\over D\right)^2\sum_{m=1} \left(1 + m Z\over m^2 \right)e^{-mZ} + {\cal O}\left(1\over D\right)^3
\end{split}
\end{equation}
Hence $\left[F(r/ r_H)-1\right]$ vanishes\footnote{ Also note that the vanishing has appropriate fall-off behavior (exponential decay in  the scaled $Z$ variable) as required by large $D$ corrections} upto corrections of order ${\cal O}\left(1\over D\right)^2$.\\
After substituting  equation \eqref{eq:Fexprep} and the value for the integration constant $\tilde C_3$, the black-brane metric dual to hydrodynamics takes the following form
\begin{equation}\label{eq:hydroset1}
\begin{split}
dS^2 = &~dS_\text{background}^2 +r^2\left(r_H\over r\right)^{D-1} (u_\mu ~ dx^\mu)^2 +{\cal O}\left(1\over D\right)^2\\
\end{split}
\end{equation}
where $dS_\text{background}^2$ is given by equation \eqref{eq:metback}

As we have mentioned before, the metric in \cite{arbBack} is described in terms of one auxiliary function $\psi$ and one auxiliary null one-form $O_A dX^A$. For convenience we are quoting the metric here again.
\begin{equation}\label{eq:quote}
dS^2 = dS_\text{background}^2 + \psi^{-D}\left(O_A ~dX^A\right)^2 + {\cal O}\left(1\over D\right)^2
\end{equation}
Here $\psi^{-D}$ is harmonic with respect to the background with $\psi =1$ being the event horizon of the full space-time and $O_A$ is simply proportional to $\bar O_A$ determined in the previous subsection. The proportionality factor  (let us denote it by the scalar function $\Phi(X)$)  is fixed using the condition that the component of $O_A $ along the unit normal of $\psi =\text{constant}$ hypersurfaces is one everywhere.
In terms of equations, the above conditions could be expressed as
\begin{equation}\label{eq:normalO}
\bar O^A = \Phi(X)~ O^A,~~~\Phi(X) = {\bar O^A ~\partial_A\psi\over \sqrt{(\partial_A\psi)(\partial^A\psi)}}~~\text{ where} ~~\partial^A\psi \equiv\bar{\cal G}^{AB}~\partial_B\psi
\end{equation}
Rewriting  \eqref{eq:quote} in terms of $\bar O_A$,
\begin{equation}\label{eq:quote2}
\begin{split}
dS^2 =~& dS_\text{background}^2 +\left( \psi^{-D}\over \Phi^2\right)\left(\bar O_A ~dX^A\right)^2 + {\cal O}\left(1\over D\right)^2\\
 =~& dS_\text{background}^2 +\left( \psi^{-D}\over \Phi^2\right)\left(u_\mu -A_2 ~a_\mu\right)\left(u_\nu -A_2 ~a_\nu\right) ~dx^\mu dx^\nu + {\cal O}\left(1\over D\right)^2\\
\end{split}
\end{equation}
The metric  in \eqref{eq:quote2} will match exactly with the metric in  \eqref{eq:hydroset1}  provided we set $A_2$ to zero and identify  $\left[\Phi^2r^2\left(r_H\over r\right)^{D-1}\right]$ with the harmonic function $\psi^{-D}$ upto corrections of order $\left(1\over D\right)^2$.
 Hence in terms of equation, what we finally have to verify is the following 
\begin{equation}\label{eq:finch}
\psi^{-D} -\Phi^2r^2\left(r_H\over r\right)^{D-1} = {\cal O}\left(1\over D\right)^2
\end{equation}
where  $\psi$ satisfies 
\begin{equation}\label{subsi1}
\nabla^2\psi^{-D} =0
\end{equation}
 with the boundary condition that $\psi =1$ should reduce to the horizon, i.e., the hypersurface given by $r=r_H$, in an expansion in $\left(1\over D\right)$.\\
 Now we shall  first determine $\psi$ and then $\Phi$. Note that both $\psi$ and the norm of $\partial_A\psi$ are scalar functions and it is much easier to compute them in a coordinate system where the background metric has a simple form. Therefore we shall solve the equation in the $\{\rho, y^\mu\}$ coordinate system and then transform the answer to the $\{r,x^\mu\}$ coordinates for final matching. 
First we need to know the position of the horizon in $\{Y^A\}$ coordinates since that will provide the required boundary condition for $\psi$. We know that in $\{X^A\} = \{r, x^\mu\}$ coordinates the horizon is at $r= r_H(x) + {\cal O}(\partial^2)$. Now $\{X^A\}$ and $\{Y^A\}$ coordinates are related as follows.
\begin{equation}\label{eq:explictransform}
\begin{split}
&\rho = r - {\Theta(x)\over D-2} + {\cal O}(\partial^2),\\
&y^\mu = x^\mu + {u^\mu(x)\over r} +\left({\Theta(x)\over D-2}\right)\left(u^\mu(x)\over r^2\right) +C_1 ~\Theta(x)~u^\mu(x) + C_2 ~a^\mu(x)+ {\cal O}(\partial^2)
\end{split}
\end{equation}
The inverse transformation:
\begin{equation}\label{eq:inversetransform}
\begin{split}
&r = \rho + {\Theta(y)\over D-2} + {\cal O}(\partial^2)\\
&x^\mu = y^\mu - {u^\mu(x)\over \rho}  -C_1 ~\Theta(x)~u^\mu(x) - C_2 ~a^\mu(x)+ {\cal O}(\partial^2)\\
&~~~=y^\mu - {u^\mu(y)\over \rho} + {a^\mu(y)\over \rho^2}  -C_1 ~\Theta(y)~u^\mu(y) -C_2 ~a^\mu(y)+{\cal O}(\partial^2)
\end{split}
\end{equation}
Therefore in terms of $\{Y^A\}$ coordinates the horizon is at 
\begin{equation}\label{eq:rhohor}
\begin{split}
 \rho =~& r_H\left(x^\mu\right) -\left({\Theta\over D-2} \right)+ {\cal O}\left(\partial^2\right)\\
 =~&r_H(y^\mu) -{\left(u\cdot\partial\right) r_H \over r_H}  -\left({\Theta\over D-2} \right)+ {\cal O}\left(\partial^2\right) = r_H(y^\mu) + {\cal O}\left(\partial^2\right) \\
 \end{split}
 \end{equation}
Here, for any term that is of  first order in derivative to begin with, this coordinate transformation will generate change of order ${\cal O}(\partial^2)$ and therefore negligible in our computation. In the last line we have used equation \eqref{eq:conservationeq}.\\
Once we know the position of the horizon, we could solve for $\psi$. In $\{\rho, y^\mu\}$ coordinates the expressions for $\psi$ and its norm are as follows (see  appendix (\ref{app:derivation} for derivation). 
\begin{equation}\label{eq:psinorm}
\begin{split}
&\psi 
(\rho, y^\mu) =  1 + \left(1-{1\over D}\right)\left({\rho\over r_H (y)} -1\right) + {\cal O}\left(1\over D\right)^3\\
\Rightarrow ~~~&dY^A~\partial_A\psi = \left(1-\frac{1}{D}\right)\left(d\rho\over r_H(y)\right)-\rho\left(1-\frac{1}{D}\right)\left(\frac{\partial_\mu r_H(y)}{r_H^2(y)}\right)~dy^\mu\\
\Rightarrow ~~~&\partial^A\psi ~\partial_A\psi =\left(\rho\over r_H(y)\right)^2\left(1-\frac{1}{D}\right)^2 + {\cal O}(\partial)^2
\end{split}
\end{equation}
Clearly this solution satisfies the  boundary condition that $\psi=1 \Rightarrow \rho = r_H(y) + {\cal O}(\partial^2)$.\\
Now we have to transform these quantities in $\{X^A\}$ coordinates.  We shall first transform the quantity $\bigg[{\rho\over r_H(y)}\bigg] $.
\begin{equation}\label{eq:rbyrh}
\begin{split}
{\rho\over r_H(y)} =~& {r -{\Theta\over D-2}\over r_H(x) +{ (u\cdot\partial)r_H\over r } }+ {\cal O}(\partial^2)\\
=~&\left(1\over r_H(x)\right)\left(r -{\Theta\over D-2}\right)\left(1-{ (u\cdot\partial)r_H\over r ~r_H}\right)+ {\cal O}(\partial^2)\\
=~&\left(1\over r_H(x)\right)\left(r -{\Theta\over D-2}-{ (u\cdot\partial)r_H\over r_H}\right)+ {\cal O}(\partial^2)= {r\over r_H(x)} +  {\cal O}(\partial^2)
\end{split}
\end{equation}
From equation \eqref{eq:rbyrh} it follows that
\begin{equation}\label{eq:psinormX}
\begin{split}
&\psi 
(r,x^\mu) =  1 + \left(1-{1\over D}\right)\left({r\over r_H (x)} -1 \right) + {\cal O}\left({1\over D^3},~\partial^2\right) \\
\Rightarrow ~~~&dX^A~\partial_A\psi = \left(1-\frac{1}{D}\right)\left(d r\over r_H\right)-r\left(1-\frac{1}{D}\right)\left(\frac{\partial_\mu r_H}{r_H^2}\right)~dx^\mu + {\cal O}\left({1\over D^2},~\partial^2\right) \\
\Rightarrow ~~~&\partial^A\psi ~\partial_A\psi =\left(r\over r_H\right)^2\left(1-\frac{1}{D}\right)^2 + {\cal O}\left({1\over D^2},~\partial^2\right) 
\end{split}
\end{equation}
Substituting this solution in equation \eqref{eq:normalO} we find
$~~~\Phi(X) = {1\over r}$.\\

Now we have all the ingredients to verify equation \eqref{eq:finch}.
 Let us introduce a new ${\cal O}(1)$ variable $R$ such that
$${r\over r_H} = 1+ {R\over D}$$
In terms of $R$ we find
\begin{equation}\label{eq:finstep}
\begin{split}
\psi^{-D} -\Phi^2r^2\left(r_H\over r\right)^{D-1} =~&\psi^{-D} - \left(r\over r_H\right)^{-(D-1)}\\
 =~& \left[1 + \left(1-{1\over D}\right)\left(R\over D\right)\right]^{-D}-\left( 1 +{R\over D}\right)^{-(D-1)}\\
 =~& -{1\over 2}\left(R\over D\right)^2 e^{-R} + {\cal O}\left(1\over D\right)^3
\end{split}
\end{equation}
This is exactly what is required to have a match between the  `hydrodynamic metric' and the `large-$D$'  metric upto the expected order.

\subsection{Comparison between the evolution of two sets of data }
As mentioned before, the `hydrodynamic metric' is defined in terms of the velocity and the  temperature \footnote{The temperature and the horizon radius are related by the following relation $$ r_H = {4\pi~ T\over (D-1)}$$
 In our choice of units $$r_H \sim {\cal O}(1)~~\Rightarrow~~T\sim {\cal O}(D)$$} of the relativistic conformal fluid moving in a flat Minkowski space-time of dimension $(D-1)$.   In case of large - $D$ expansion, the metric is given in terms of a $(D-1)$ dimensional time-like fluctuating membrane embedded in pure AdS space-time with a dynamical velocity field on it.  Both of these two sets of data are controlled by separate equations. For `derivative expansion' , the governing equation of data is given in \eqref{eq:conservationeq}. In `large-$D$' technique, the relevant equation is the following\cite{arbBack}
 \begin{equation}\label{eq:largeDeqn}
 \begin{split}
 &\hat\nabla\cdot U=0,~~~~~
 \left[\frac{\hat{\nabla}^2 U_\alpha}{\cal{K}}-\frac{\hat{\nabla}_\alpha{\cal{K}}}{\cal{K}}+U^\beta {\cal{K}}_{\beta\alpha}-U\cdot\hat{\nabla} U_\alpha\right]{\cal P}^\alpha_\gamma=0
 \end{split}
 \end{equation}
 Here  the equation is written as an intrinsic equation on the membrane world-volume. All raising, lowering and contraction of the indices are done with respect to the induced metric on the dynamical membrane. $U_\alpha$ is the velocity of the membrane, expressed in terms of its intrinsic coordinates. ${\cal{K}}_{\beta\alpha}$ is the extrinsic curvature of the membrane, expressed as a symmetric tensor on the membrane world-volume.  ${\cal{K}}$ denotes its trace. ${\cal P}^\alpha_\gamma$ is the projector perpendicular to $U^\alpha$.\\
In this subsection, our goal is to show that equation \eqref{eq:conservationeq} implies equation \eqref{eq:largeDeqn} upto corrections of order ${\cal O}\left(1\over D\right)^2$. 
 
 Our first job would be  to express  the $U^\alpha$ and ${\cal K}_{\alpha\beta}$ in terms of velocity $u^\mu$ and temperature (or $r_H$) of the relativistic fluid. Remember that though both $u^\mu$ and $U^\alpha$ are unit normalized velocity vector, they are defined on completely different spaces, one being a flat Minkowski metric and the other is the curved (both intrinsic and extrinsic curvature, being nonzero) membrane world volume.  \\
 For convenience, we shall work in  $\{Y^A\} = \{\rho, y^\mu \}$ coordinates where the background metric is simple.
  We shall first compute the unit normal to the membrane and different components of its extrinsic curvature, to begin with in terms of background coordinates and then we shall re-express it as an intrinsic symmetric tensor on the membrane.\\
 The unit normal to the membrane is given by
\begin{equation}\label{eq:unitnormal}
\begin{split}
n_A~dY^A\vert_\text{membrane} \equiv &~dY^A \left[{\partial_A\psi\over\sqrt{\partial^A\psi~\partial_A\psi}}\right]_\text{membrane}\\
&= {d\rho -dy^\mu~\partial_\mu r_H(y)\over r_H(y)}
\end{split}
\end{equation}
The extrinsic curvature is defined as follows.
\begin{equation}\label{eq:extrinsicdef}
\begin{split}
K_{AB}&=\Pi^C_A~\nabla_C n_B
=\Pi^C_A\left(\partial_C n_B-\Gamma_{CB}^D n_D\right)\\
\text{where}&~\Pi^B_A=\delta^B_A-n_A~n^B~~\text{and $\nabla$ is the covariant derivative w.r.t background}
\end{split}
\end{equation}
Now let us choose $\{y^\mu\}$ as the intrinsic coordinate on the membrane world volume. In this choice of coordinates, the extrinsic curvature ${\cal K}_{\alpha\beta}$ will have the following structure.
\begin{equation}\label{eq:extrinsicintrin}
\begin{split}
{\cal K}_{\alpha\beta}&=K_{\rho\rho} \left(\partial_\alpha r_H\right)\left(\partial_\beta r_H\right) + \left[ K_{\rho\alpha}\left(\partial_\beta r_H\right) + K_{\rho\beta}\left(\partial_\alpha r_H\right) \right] + K_{\alpha\beta}
\end{split}
\end{equation}
Note that the first term in the RHS of equation \eqref{eq:extrinsicintrin} does not contribute at first order derivative expansion.\\
After using equation \eqref{eq:extrinsicdef} and \eqref{eq:extrinsicintrin}, at this order the final expression for  ${\cal K}_{\mu\nu}$ turns out to be very simple (see appendix (\ref{app:extrinsic}) for the details of the computation).
\begin{equation}\label{eq:diffcompex}
\begin{split}
{\cal K}_{\alpha\beta}&=r_H^2~\eta_{\alpha\beta} + {\cal O}(\partial^2),~~~{\cal K} = (D-1)
\end{split}
\end{equation}
The induced metric on the membrane is given by
\begin{equation}\label{eq:diffinduc}
\begin{split}
g_{\alpha\beta}&=r_H^2~\eta_{\alpha\beta} + {\cal O}(\partial^2)
\end{split}
\end{equation}
Now we shall determine the velocity $U^\alpha$. The  velocity is defined as the projection of $O^A$ on the membrane which, by construction,  would be unit normalized with respect to the induced metric of the membrane. In $\{Y^A\}$ coordinates, $O_A~dY^A$ takes the following form
\begin{equation}\label{eq:Otransform}
\begin{split}
&O_A~dX^A\vert_\text{membrane} =-\left[r ~u_\mu (x)~ dx^\mu\right]_\text{membrane}\\
=~&-\left(r_H(y)+{\Theta\over D-2}\right)\left[u_\mu (y) -{a_\mu(y)\over r_H}\right] \left[\left({\partial x^\mu\over \partial \rho}\right)d\rho+\left({\partial x^\mu\over \partial y^\nu}\right)dy^\nu\right]_{\rho =r_H(y)}\\
=~&-\left(r_H(y)+{\Theta\over D-2}\right)\left[u_\mu (y) -{a_\mu(y)\over r_H}\right] \left[\left({u^\mu(y)\over r_H^2(y)}- {2a^\mu(y)\over r_H^3(y)}\right)d\rho+\left(\delta_\nu^\mu-\frac{\partial_\nu u^\mu}{r_H}\right)dy^\nu\right]\\
=~& \left({1\over r_H(y)}+ {\Theta \over {(D-2)r_H^2}}\right)d\rho +\left[- r_H(y)~u_\mu(y) -\left(\Theta\over D-2\right) u_\mu +a_\mu(y)\right]~dy^\mu\\
=~& \left({1\over r_H(y)}+ {\Theta \over {(D-2)r_H^2}}\right)d\rho +\left[- r_H(y)~u_\mu(y) -\left(\partial_\mu r_H\over r_H\right)\right]~dy^\mu
\end{split}
\end{equation}
In the last line we have used equation \eqref{eq:conservationeq}, which is the governing equation for the data in the hydrodynamic side of the duality.

From equations \eqref{eq:Otransform} and \eqref{eq:unitnormal} it follows that 
\begin{equation}
\begin{split}
U_A~dY^A\equiv &-dY^A \left[O_A - n_A\right]_\text{membrane} =-\left(1\over r_H^2\right) \left(\Theta\over D-2\right) d\rho + r_H~u_\mu ~ dy^\mu
\end{split}
\end{equation}
Now $U_\alpha$ is just rewriting of $U_A$ in terms of the intrinsic coordinates of the membrane. Following the same method as in equation \eqref{eq:extrinsicintrin} we find 

\begin{equation}\label{eq:Umu}
\begin{split}
&U_\alpha~dy^\alpha\equiv \left[r_H ~u_\alpha +  {\cal O}(\partial^2)\right] ~dy^\alpha\\
\end{split}
\end{equation}

Once we know ${\cal K}_{\alpha\beta}$, $U^\alpha$ and the induced metric on the membrane, we could compute each term in the equation  \eqref{eq:largeDeqn}.
\begin{equation}\label{eq:eachterm}
\begin{split}
&\hat\nabla\cdot U =\left(D-2\over r_H\right)\left[{\Theta\over D-2} + {(u\cdot\partial)r_H\over r_H}\right] + {\cal O}\left(\partial^2\right)= {\cal O}\left(\partial^2\right)\\
&\hat\nabla^2 U_\alpha = {\cal O}\left(\partial^2\right)\\
&(U\cdot\hat\nabla)U_\beta = a_\beta + {P^\alpha_\beta\partial_\alpha~ r_H\over r_H} +  {\cal O}\left(\partial^2\right) ={\cal O}\left(\partial^2\right)\\
&U^\alpha~{\cal K}_{\alpha\beta} ~{\cal P}^\beta_\gamma = {\cal O}(\partial^2)\\
&\hat\nabla_\alpha{\cal K} = {\cal O}(\partial^2)
\end{split}
\end{equation} 
As it is clear from the notation, in the LHS of each equation the relevant metric is the induced metric on the membrane whereas in RHS it is the flat Minkowski metric $\eta_{\alpha\beta}$.\\
Substituting equations \eqref{eq:eachterm} in equation \eqref{eq:largeDeqn} we could easily show that membrane equation follows as a consequence of fluid equation.

In this context let us mention the work in \cite{Dandekar:2017aiv}. Here the authors have computed the boundary  stress tensor dual to a slowly varying membrane embedded in AdS. They have found the dual fluid velocity in terms of the membrane velocity. It could be easily checked that equation \eqref{eq:Umu} is indeed the inverse of what they have found upto correction of order ${\cal O}(\partial^2)$.

\section{Conclusion}\label{sec:conclude}
In this note we have compared dynamical black-brane solutions to Einstein's equations (in presence of negative cosmological constant) generated by two different perturbative schemes, namely `derivative expansion' and Large-dimension expansion. In both the cases, the space-time necessarily have an event horizon.  We have shown that in large number of dimensions whenever `derivative expansion' is applicable, we can expand the metric further in $\left(1\over D\right)$, (though the reverse may not be true always). We have found perfect match in this overlap regime of these two perturbative techniques upto first subleading order on both sides. \\
One immediate interesting project would be to extend this calculation to the next  order on both sides, since we already know both the `hydrodynamic  metric' and the `large D metric' upto the second subleading order \cite{nonlinfluid,secondorder}. It would also be interesting to generalize this calculation to Einstein-Maxwell system in presence of negative cosmological constant, where also we know the metric on both sides upto the first subleading order\cite{Banerjee:2008th,Chmembrane,Erdmenger:2008rm,chargepoulomi}.\\

In some sense, our analysis serves as a consistency test for these two methods. But this comparison could teach us something more. This  is about the dual systems of these two gravity solutions. \\
The dynamical black-brane metric generated by `derivative expansion' in $D$ dimension is dual to the relativistic conformal hydrodynamics living in $(D-1)$ dimensional flat space-time. The variables of hydrodynamics are fluid velocity and temperature, which are the data that label different black-brane solutions in derivative expansion. 
\\
On the other hand the metric generated in `large $D$ expansion' is dual to a  co-dimension one dynamical membrane embedded in pure AdS and coupled with a velocity field.  Here also the labeling data of the metric live on a $(D-1)$ dimensional hypersurface and they consist of a scalar function - the shape of the membrane and a unit normalized velocity field. This is very similar to hydrodynamics in terms of counting, though the governing equations and the physical significance of the variables are entirely different. 

However, we have already seen that these two systems of equations are approximately equivalent after an appropriate field redefinition. In this note, we have verified it at the very leading order  and we expect that the project of comparing the two metric upto second subleading order would extend  this equivalence to the next order on both sides.

In fact it is  expected that this equivalence is  valid to all orders\cite{Dandekar:2017aiv}. In other words, in the overlap regime, these two equations must be exactly equivalent to each other if we consider all orders on both sides\cite{Dandekar:2017aiv}, though to see this equivalence we need to re-express the variables of one side in terms of the other \cite{Dandekar:2017aiv,AmosTurbulence,Andrade:2018zeb}.\\
This equivalence actually involves some interesting resum of one series into the other. Even the leading term in derivative expansion can encode many terms of $\left(1\over D\right)$ expansion and on the other hand the leading membrane equation might have information about many higher order transport coefficientas.  At linearlized level, this has been nicely captured in the analysis in \cite{EmparanHydro}. The frequencies of Quasi normal modes do exhibit such resum.
In \cite{Dandekar:2017aiv}, the authors have proposed a resummed stress tensor that could exactly reproduce the fluid stress tensor exactly upto the first order in derivative expansion.
It would be very interesting to understand this structure in full detail, at non linear level. This might lead to a fluid-membrane duality in large number of dimensions where gravity does not have any role to play.

\section*{Acknowledgment}
It is a great pleasure to thank Shiraz Minwalla for initiating discussions on this topic and for his numerous suggestions throughout the course of this work.   We would  like to thank Yogesh Dandekar for illuminating discussions. 
We would like to thank Yogesh Dandekar, Shiraz Minwalla and Yogesh Kumar Srivastava for reading through the initial draft and very useful comments.\\
 We  would  also like to acknowledge our debt to the people of India for their steady and generous support to research in the basic sciences.

\appendix
\section{Analysis of $F(r/r_H)$}\label{app:integral}
In this section we shall evaluate the integral \eqref{eq:integral} in large $D$ limit.
For convenience we are quoting  the equation here.
\begin{equation}\label{eq:integralrep}
 F(y)=y\int_{y}^{\infty}dx\frac{x^{D-2}-1}{x(x^{D-1}-1)}
\end{equation}
We would like to evaluate this integral systematically for large $D$.
Let us first evaluate the integral for $y\geq 2$. In this case, since $D$ is very large, $x^D>>1$ throughout the range of integration. So we shall expand the integrand in the following way.

\begin{equation}\label{eq:expint}
\begin{split}
\frac{x^{D-2}-1}{x(x^{D-1}-1)} =&~ \left(1\over x^2\right)\left(1-x^{-(D-2)}\right)\left(1-x^{-(D-1)}\right)^{-1}\\
=&~\left(1\over x^2\right)\left(1-x^{-(D-2)}\right)\left(1+\sum_{m=1}x^{-m(D-1)}\right)\\
=&~\left(1\over x^2\right)\bigg(1+\sum_{m=1}\left[x^{-m(D-1)} - x^{-m(D-1) +1}\right]\bigg)
\end{split}
\end{equation}
Integrating \eqref{eq:expint} we find
\begin{equation}\label{eq:inteser}
\begin{split}
y\int_{y\geq2}^{\infty}dx\frac{x^{D-2}-1}{x(x^{D-1}-1)} = 1 +\sum_{m=1} \bigg[\left(1\over  (D-1)m  +1\right)y^{-(D-1)m } - \left(1\over (D-1)m \right)y^{-(D-1)m+1 } \bigg]
\end{split}
\end{equation}
Clearly the sums in the RHS of \eqref{eq:inteser} are convergent for $y\geq2$. Let us denote the RHS as $k(y)$.\\
However, the expansion in \eqref{eq:expint} is not valid inside the `membrane region', i.e.,  when $y-1\sim {\cal O}\left(1\over D\right)$ and naively $k(y)$ is not the answer for the integral.

But consider the function $\tilde k(y)=F(y) -k(y)$. This function vanishes for all $y\geq2$ and also by construction it is a smooth function at $y=2$ (none of the derivatives diverge). Hence $\tilde k(y)$ must vanish for every $y$.
So we conclude, for every allowed $y$ (i.e., $y\geq1$)
\begin{equation}\label{eq:finint}
\begin{split}
F(y)= 1 +\sum_{m=1} \bigg[\left(1\over  (D-1)m  +1\right)y^{-(D-1)m } - \left(1\over (D-1)m \right)y^{-(D-1)m+1 } \bigg]
\end{split}
\end{equation}
Note that $F(y)$ reduces to $1$ as $y\rightarrow\infty$ as required in section (\ref{subsec:1stderi}).\\
Now we would like to expand $F(y)$ in a series in $\left(1\over D\right)$, where $y$ is in the membrane regime.
$$y = 1 +{Y\over D },~~~Y\sim {\cal O}(1)$$
In this regime $F(y)$ takes the following form
\begin{equation}\label{eq:Fexp}
\begin{split}
F(y) = F\left(1 +{Y\over D}\right) = 1 - \left(1\over D\right)^2\sum_{m=1} \left(1 + m Y\over m^2 \right)e^{-mY} + {\cal O}\left(1\over D^3\right)
\end{split}
\end{equation}
In this note we considering only the first subleading correction in $\left(1\over D\right)$ expansion. Therefore $F(y)$ could be set to 1 for our purpose.

\section{Derivation of $\psi$ in $\{Y^A\} = \{\rho, y^\mu\}$ coordinates}\label{app:derivation}
In this section we shall give the derivation of $\psi$ as mentioned in eq \eqref{subsi1}. We want to solve $\psi$ such that $\nabla^2\psi^{-D}=0$. Where $\nabla$ is the covariant derivative with respect to the background metric
\begin{equation}
{ds}^{2}_{background}=\frac{{d\rho}^{2}}{{\rho}^{2}}+{\rho}^{2}\eta_{\mu\nu} ~dy^\mu ~dy^\nu
\end{equation}
we can expand $\psi$ as follows
\begin{equation}
\psi=1+\left(A_{10}+\epsilon~B_{10}+\frac{A_{11}+\epsilon~B_{11}}{D}\right)(\rho-r_H)+(A_{20}+\epsilon~B_{20})(\rho-r_H)^2+{\cal O}\left(\frac{1}{D^3}\right)
\end{equation}
Here $\epsilon$ denotes that $B_{ij}$'s are ${\cal O}(\partial)$ terms.
\begin{equation}\label{del2psiD}
\begin{split}
&~~~~~~\nabla^2\left(\psi^{-D}\right)=0\\
&\Rightarrow \psi \left(\nabla^2 \psi\right)-(D+1)(\nabla^A \psi)(\nabla_A \psi)=0\\
&\Rightarrow \psi~\rho^2\bigg[\partial_\rho\partial_\rho \psi-\Gamma^\rho_{\rho\rho}(\partial_\rho\psi)-\Gamma^\mu_{\rho\rho}(\partial_\mu \psi)\bigg]+\frac{\psi}{\rho^2}~\eta^{\mu\nu}\bigg[-\Gamma^\rho_{\mu\nu}(\partial_\rho\psi)-\Gamma^\alpha_{\mu\nu}\partial_\alpha\psi\bigg]\\
&~~~~~~~~~~~~~~~~-(D+1)~\rho^2~(\partial_\rho\psi)^2+{\cal O}(\partial)^2=0
\end{split}
\end{equation}
The required Christoffel symbols are
\begin{equation}
\begin{split}
\Gamma^\rho_{\rho\rho}=-\frac{1}{\rho};~~~~~\Gamma^\mu_{\rho\rho}=0;~~~~~\Gamma^\rho_{\mu\nu}=-\rho^3\eta_{\mu\nu};~~~~~\Gamma^\alpha_{\mu\nu}=0;
\end{split}
\end{equation}
Using the above Christoffel symbol we get
\begin{equation}\label{psid}
\psi\bigg[\rho^2~\partial_\rho^2 \psi+D\rho~\partial_\rho\psi\bigg]-(D+1)~\rho^2~(\partial_\rho\psi)^2=0
\end{equation}
Now,
\begin{equation}\label{eq:delrhopsi}
\begin{split}
\partial_\rho \psi&=\left(A_{10}+\epsilon~B_{10}+\frac{A_{11}+\epsilon~B_{11}}{D}\right)+2~(A_{20}+\epsilon~B_{20})(\rho-r_H)\\
\partial_\rho^2 \psi&=2~(A_{20}+\epsilon~B_{20})
\end{split}
\end{equation}\\
Solving, \eqref{psid} order by order in derivative expansion we get the following solution
\begin{equation}
\psi(\rho,y^\mu)=1+\left(1-\frac{1}{D}\right)\left(\frac{\rho}{r_H(y^\mu)}-1\right)+{\cal O}\left(\frac{1}{D}\right)^3
\end{equation}
\section{Computing different terms in membrane equation}\label{app:extrinsic}
In this section we shall give the details of calculations of different terms that appear in the membrane equation. The different components of the projector defined in \eqref{eq:extrinsicdef} are given by
\begin{equation}
\begin{split}
&\Pi^\rho_\rho=0;~~~~~\Pi^\rho_\mu=\partial_\mu r_H;~~~~~\Pi^\mu_\rho=\frac{1}{r_H^4}(\partial^\mu r_H);~~~~~\Pi^\mu_\nu=\delta^\mu_\nu
\end{split}
\end{equation}
The different components of Christoffel symbol of the background metric in $Y^A=\{\rho,y^\mu\}$ co-ordinates are given by
\begin{equation}
\begin{split}
\Gamma^\rho_{\rho\rho}=-\frac{1}{\rho};~~~~~\Gamma^\rho_{\mu\rho}=0;~~~~~\Gamma^\rho_{\mu\nu}=-\rho^3\eta_{\mu\nu};~~~~~\Gamma^\nu_{\mu\rho}=\frac{1}{\rho}\delta^\nu_\mu;~~~~~\Gamma^\alpha_{\mu\nu}=0;~~~~~\Gamma^\mu_{\rho\rho}=0;
\end{split}
\end{equation}\\
From \eqref{eq:extrinsicintrin} it is clear that we need only $K_{\rho\alpha}$ and $K_{\alpha\beta}$ component of extrinsic curvature
\begin{equation}
\begin{split}
K_{\rho\mu}&=\Pi^C_\rho\bigg(\partial_C n_\mu-\Gamma_{C\mu}^D n_D\bigg)\\
&=\Pi^\nu_\rho\bigg(\partial_\nu n_\mu-\Gamma_{\nu\mu}^\rho n_\rho\bigg)\\
&=\frac{\partial_\mu r_H}{r_H^2}\\ \\
K_{\mu\nu}&=\Pi^C_\mu\bigg(\partial_C n_\nu-\Gamma_{C\nu}^D n_D\bigg)\\
&=\Pi^\rho_\mu\bigg(\partial_\rho n_\nu-\Gamma_{\rho\nu}^\rho n_\rho\bigg)+\Pi^\alpha_\mu\bigg(\partial_\alpha n_\nu-\Gamma_{\alpha\nu}^\rho n_\rho\bigg)\\
&=-\delta^\alpha_\mu~\Gamma^\rho_{\alpha\nu}n_\rho\\
&=\rho^2~ \eta_{\mu\nu}
\end{split}
\end{equation}
Now, as mentioned in \eqref{eq:extrinsicintrin} in terms of the intrinsic coordinates on the membrane the extrinsic curvature will have the structure
\begin{equation}
\begin{split}
{\cal K}_{\alpha\beta}&=K_{\rho\rho} \left(\partial_\alpha r_H\right)\left(\partial_\beta r_H\right) + \left[ K_{\rho\alpha}\left(\partial_\beta r_H\right) + K_{\rho\beta}\left(\partial_\alpha r_H\right) \right] + K_{\alpha\beta}\\
&=r_H^2~\eta_{\alpha\beta}+{\cal O}(\partial)^2
\end{split}
\end{equation}
The trace of the extrinsic curvature
\begin{equation}
{\cal K}=(D-1)+{\cal O}(\partial^2)
\end{equation}
For the calculation of only the extrinsic curvature we need background metric, where for the rest of the calculation we require induced metric on the horizon. The induced metric on the horizon is given by
\begin{equation}
g_{\alpha\beta}=r_H^2~\eta_{\alpha\beta} + {\cal O}(\partial^2)
\end{equation}
The Christoffel symbol of the induced metric
\begin{equation}
\begin{split}
\Gamma^\delta_{\beta\alpha}=\bigg(\delta^\delta_\beta\frac{\partial_\alpha r_H}{r_H}+\delta^\delta_\alpha\frac{\partial_\beta r_H}{r_H}-\eta_{\alpha\beta}\frac{\partial^\delta r_H}{r_H}\bigg)
\end{split}
\end{equation}
Now we shall calculate all the terms mentioned in \eqref{eq:eachterm}. First we shall calculate
\begin{equation}
\begin{split}
\hat{\nabla}\cdot U&=g^{\alpha\beta}~\hat{\nabla}_\alpha U_\beta\\
&=\frac{\eta^{\alpha\beta}}{r_H^2}\left[\partial_\alpha U_\beta-\Gamma^\delta_{\alpha\beta}U_\delta\right]+{\cal O}(\partial)^2\\
&= \frac{\eta^{\alpha\beta}}{r_H^2}\left[\partial_\alpha\left(r_H~u_\beta\right)-(r_H~u_\delta)\bigg(\delta^\delta_\beta\frac{\partial_\alpha r_H}{r_H}+\delta^\delta_\alpha\frac{\partial_\beta r_H}{r_H}-\eta_{\alpha\beta}\frac{\partial^\delta r_H}{r_H}\bigg)\right]+{\cal O}(\partial)^2\\
&=(D-2)\left(\frac{(u\cdot\partial)r_H}{r_H^2}\right)+\frac{\partial\cdot u}{r_H}+{\cal O}(\partial)^2
\end{split}
\end{equation}
Now we shall calculate $\hat{\nabla}^2 U_\mu$ and $\left(U\cdot\hat{\nabla}\right)U_\alpha$
\begin{equation}
\begin{split}
\hat{\nabla}^2 U_\mu&=g^{\alpha\beta}\hat{\nabla}_\alpha\hat{\nabla}_\beta U_\mu\\
&=g^{\alpha\beta}\left[\partial_\alpha(\hat{\nabla}_\beta U_\mu)-\Gamma^\delta_{\alpha\beta}(\hat{\nabla}_\delta U_\mu)-\Gamma^\delta_{\alpha\mu}(\hat{\nabla}_\beta U_\delta)\right]\\
&={\cal O}(\partial)^2
\end{split}
\end{equation}

\begin{equation}
\begin{split}
\left(U\cdot\hat{\nabla}\right)U_\alpha&=U^\beta(\partial_\beta U_\alpha)-U^\beta~\Gamma^\delta_{\beta\alpha}U_\delta\\
&=\frac{u^\beta}{r_H}\bigg(r_H(\partial_\beta u_\alpha)+u_\alpha(\partial_\beta r_H)\bigg)-\frac{u^\beta}{r_H}(r_H~u_\delta)\bigg(\delta^\delta_\beta\frac{\partial_\alpha r_H}{r_H}+\delta^\delta_\alpha\frac{\partial_\beta r_H}{r_H}-\eta_{\alpha\beta}\frac{\partial^\delta r_H}{r_H}\bigg)+{\cal O}(\partial^2)\\
&=(u\cdot\partial)u_\alpha+u_\alpha\left(\frac{(u\cdot\partial)r_H}{r_H}\right)+\frac{\partial_\alpha r_H}{r_H}+{\cal O}(\partial^2)\\
\end{split}
\end{equation}\\
Now,
\begin{equation}
\begin{split}
U^\alpha~ {\cal K}_{\alpha\beta}~{\cal P}^\beta_\gamma&=(\delta^\beta_\gamma+U^\beta~U_\gamma)(U^\alpha~r_H^2~\eta_{\alpha\beta})+{\cal O}(\partial^2)\\
&=(\delta^\beta_\gamma+U^\beta~U_\gamma)U_\beta+{\cal O}(\partial^2)\\
&={\cal O}(\partial^2)
\end{split}
\end{equation}

\bibliographystyle{JHEP}
\bibliography{larged}

\end{document}